\documentclass[a4paper,fleqn,usenatbib]{mnras}

\usepackage{newtxtext,newtxmath}

\usepackage[T1]{fontenc}
\usepackage{ae,aecompl}

\usepackage{graphicx}
\usepackage{deluxetable}
\usepackage{lscape}
\usepackage{epsfig}
\usepackage{amssymb,amsmath}
\newcommand{\HI}{\mbox{H\,{\sc i}}}
\newcommand{\MgII}{\mbox{Mg\,{\sc ii}}}
\newcommand{\MgI}{\mbox{Mg\,{\sc i}}}
\newcommand{\FeII}{\mbox{Fe\,{\sc ii}}}

\newcommand {\apgt} {\ {\raise-.5ex\hbox{$\buildrel>\over\sim$}}\ }
\newcommand {\aplt} {\ {\raise-.5ex\hbox{$\buildrel<\over\sim$}}\ }

\newcommand{\Wmi}{$W_0^{\lambda2796}$}
\newcommand{\Wf}{$W_0^{\lambda2600}$}

\title[DLA statistics at $z<1.65$]{The Statistical Properties of Neutral Gas at $z<1.65$ from UV Measurements of Damped Lyman Alpha Systems\thanks{Based on data obtained from the Sloan Digital Sky Survey (SDSS) and on observations made with the {\it Hubble Space Telescope} (HST) operated by STScI-AURA for NASA/ESA and the NASA {\it Galaxy Evolution Explorer} (GALEX) operated for NASA by the California Institute of Technology under NASA contract NAS5-98034.}}

\author[S. Rao  et al.]{
Sandhya M. Rao,$^{1}$\thanks{E-mail srao@pitt.edu}
David A. Turnshek,$^{1}$
Gendith M. Sardane$^{1}$
and
Eric M. Monier$^{2}$
\\
$^{1}$Department of Physics and
Astronomy and PITTsburgh Particle physics, Astrophysics, and Cosmology Center 
(PITT PACC),\\
 University of Pittsburgh, Pittsburgh, PA 15260, USA\\
$^{2}$Department of Physics, The College at Brockport, State University
of New York, Brockport, NY 14420, USA\\
}

\date{Accepted XXX. Received YYY; in original form ZZZ}

\pubyear{2017}

\begin{document}
\label{firstpage}
\pagerange{\pageref{firstpage}--\pageref{lastpage}}
\maketitle

\begin{abstract}

We derive the statistical properties of neutral gas at redshifts $0.11<z<1.65$ from UV measurements of quasar Ly$\alpha$ absorption lines corresponding to 369 \MgII\ systems with $W^{\lambda2796}_{0} \ge 0.3$ \AA. In addition to the 41 damped Lyman alpha (DLA) systems presented in Rao et al. (2006), the current DLA sample includes 29 newly discovered DLAs. Of these, 26 were found in our {\it Hubble Space Telescope} (HST) Advanced Camera for Surveys prism survey for DLAs (Turnshek et al. 2015) and three were found in a {\it GALaxy Evolution Explorer} (GALEX) archival search. In addition, an HST Cosmic Origins Spectrograph Cycle 19 survey yielded no DLAs that could be used for this study. Formally, this DLA sample includes 70 systems with $N_{\rm HI}\ge 2\times 10^{20}$ atoms cm$^{-2}$. We find that the incidence of DLAs, or the product of their gas cross section and their comoving number density, can be described by $n_{\rm DLA}(z) = (0.027 \pm 0.007) (1+z)^{(1.682 \pm 0.200)}$ over the redshift range $0<z<5$. The cosmic mass density of neutral gas can be described by $\Omega_{\rm DLA}(z) = (4.77 \pm 1.60)\times10^{-4} (1 + z)^{(0.64\pm 0.27)}$. The low-redshift column density distribution function is well-fitted by a power law of the form $f(N) \sim N^\beta$ with $\beta = -1.46 \pm 0.20$. It is consistent with the high-redshift as well as $z=0$ estimates at the high column density end but, lies between them at the low column density end. We discuss possible $N_{\rm HI}$ and metallicity bias in \MgII-selected DLA samples and show that such biases do not exist in the current data at $z<1.65$. Thus, at least at $z<1.65$, DLAs found through \MgII\ selection statistically represent the true population of DLAs. However, we caution that studies of DLA metallicities should take into the account the relative incidence of DLAs with respect to \Wmi\ (or gas velocity spread) in order to correctly measure the mean neutral-gas cosmic metallicity of the universe.

\end{abstract}


\section{Introduction}

For decades, quasar absorption line surveys have proved to be a powerful and highly successful way of probing intervening gaseous structures in the Universe. Studying the redshift evolution of absorption lines, be they lines of molecular H$_2$, neutral \HI, or metals, allows us to trace the star formation and galaxy assembly history of the Universe without being biased by the luminous components of galaxies. In particular, the damped Lyman alpha (DLA) systems, which have the highest observed neutral hydrogen column densities, $N_{\rm HI} \ge 2 \times 10^{20}$ atoms cm$^{-2}$, are known to contain the bulk of the neutral gas mass in the Universe (see Wolfe et al.  2005 for an overview of DLA research.) Since the original Lick survey for DLAs (Wolfe et al. 1986), many teams have advanced the field at redshifts $z\ge 1.65$, where the 
Ly$\alpha$ line falls in the optical part of the spectrum and can be accessed with ground-based telescopes (Turnshek et al. 1989; Lanzetta et al. 1991; Storrie-Lombardi \& Wolfe 2000; P\'eroux et al. 2003; Prochaska \& Herbert-Fort 2004;  Prochaska, Herbert-Fort, \& Wolfe 2005; Ellison et al. 2008; Prochaska \& Wolfe 2009; Noterdaeme et al. 2009; 2012; Zafar et al. 2013a; 2013b; Crighton et al. 2015; S\'anchez-Ram\'irez et al. 2016). This was accomplished by measuring the properties of the DLAs and considering successively larger samples, which statistically improved the accuracy of the results on the DLA incidence and cosmic neutral gas mass density at $z\ge 1.65$. The number of known DLA systems at these higher redshifts increased from $\sim 15$ in the Lick survey to over 6800 in the Sloan Digital Sky Survey III (Data Release 9) quasar spectroscopic database (Noterdaeme et al. 2012, henceforth, N12). 

Nevertheless, DLAs are rare, especially at redshifts $z<1.65$ -- an interval that includes the most recent $\sim$70\% of the age of the Universe -- where the line falls in the UV. The HST Quasar Absorption Line Key Project found only 1 DLA in a blind survey (Bahcall et al. 1993), while Meiring et al. (2011) found 3 serendipitous DLAs in the COS-Halos survey, and Neeleman et al. (2016) identified 4 DLAs in a blind survey of 463 quasars in the HST archives. These results are consistent given the small samples and low statistical accuracies. Thus, given the limited availability of space-based observing time for UV surveys, the identification of a substantial sample of `UV'-redshift DLAs in a blind spectroscopic survey is practically impossible, not to mention extremely inefficient. Therefore, to isolate sightlines which have DLAs, we used a technique that exploited the fact that all known optical-redshift DLAs have strong \MgII\ absorption, i.e., with \MgII$\lambda 2796$ rest equivalent width $W_0^{\lambda2796}\ge 0.3$  \AA, to construct a survey for DLAs in a sample of \MgII\ absorbers (Rao, Turnshek, \& Briggs 1995; Rao \& Turnshek 2000; Rao, Turnshek, \& Nestor 2006, henceforth RTN06). We emphasize here that the strong-\MgII-selection method is primarily a selection based on gas velocity spread since \MgII\ absorption lines are generally saturated. Therefore, the lines become strong when their velocity spread (or number of components) is large, and not because the gas has high metallicity. This is especially true at the SDSS spectral resolution, where individual components of \MgII\ absorption lines usually cannot be resolved. We will later show clear evidence that refutes any indication of a metallicity bias at $z<1.65$.   
 
In this paper, we present a compilation of low-redshift DLAs that is statistically complete for the purpose of determining neutral gas properties at $z<1.65$.  We extend the RTN06 UV-DLA sample with three additional \MgII-selected surveys. The first is our HST ACS-HRC-PR200L prism survey for DLAs in the redshift interval $0.42<z<0.70$, in which we found 35 high-probability DLA systems, of which 26 formally have $N_{\rm HI} \ge 2 \times 10^{20}$ atoms cm$^{-2}$ (Turnshek et al. 2015). The second is a {\it GALaxy Evolution Explorer} (GALEX) archival survey in which we found 3 DLAs (this paper). The third is an HST Cycle 19 survey of 16 $z<0.4$ \MgII\ systems that were selected from our MMT survey for \MgII\ systems (Nestor et al. 2006 and this paper), none of which are DLAs. In all, we now have 369 \MgII\ systems at $z<1.65$ for which UV spectra reveal the nature of the corresponding Ly$\alpha$ line; 70 of these are DLAs. Without the possibility of being able to obtain large samples of UV spectra of quasars in the near future, these results will define the state of the field for some time to come. 

We describe the three surveys in more detail in \S 2. The statistical properties of neutral gas at $z<1.65$ as determined from the updated DLA sample are presented in \S 3. Potential biases are discussed in \S 4, and we conclude with a summary in \S 5. Throughout we assume a ``737'' cosmology with H$_0$ = 70 km s$^{-1}$ Mpc$^{-1}$, $\Omega_{M} = 0.3$, and $\Omega_{\Lambda} = 0.7$.

\section{The UV-DLA Surveys}

In RTN06 we presented results from searches for DLAs in UV spectra of quasars with low-redshift ($z<1.65$) \MgII\ systems. This started as a compilation of \MgII\ systems from various sources that included those from the literature for which  {\it International Ultraviolet Explorer} (IUE) as well as early HST FOS data were available in the archives (Rao et al. 1995). To this dataset we added \MgII-selected DLA survey results from programs in HST Cycle 6 (Rao \& Turnshek 2000) and Cycles 9 and 11 (RTN06), as well as additional results that could be culled from the literature using our SDSS Early Data Release quasar \MgII\ survey (Nestor et al. 2005, henceforth, NTR05). In all, the RTN06 sample included 41 DLAs found among 197 \MgII\ systems with $W_0^{\lambda2796} \ge 0.3$. All the DLAs in the sample had $W_0^{\lambda2796} \ge 0.6$ \AA\ and 50\% of systems with \MgII\ $W_0^{\lambda2796}$ and \FeII\ $W_0^{\lambda2600} > 0.5$ \AA\ were DLAs. We note that in the expanded sample presented here, only one of the 70 DLAs has $W_0^{\lambda2796} < 0.6 $\AA.

With the 41 DLAs in our \MgII-follow-up sample (RTN06), we showed that the incidence of DLAs, i.e., $dn/dz$ or the number of DLAs per unit redshift, which is a product of their volume number density and physical cross-section on the sky, showed no evolution for redshifts $z<2$, but evolved significantly in comparison to the high-redshift DLA results available at the time (Prochaska \& Herbert-Fort 2004). On the other hand, the cosmological mass density of neutral gas as traced by the DLAs, $\Omega_{\rm DLA}$, remained constant to within the statisitical errors for redshifts $0.5<z<5$. Our sample also had a higher fraction of high-column density DLAs ($\log N_{\rm HI} >21.5$ cm$^{-2}$), which made the low-redshift column density distribution, $f(N)$, flatter at higher column densities in comparison to the distributions at high redshift and $z=0$ (Zwaan et al. 2005, henceforth, Z05).

Since then, the $z>1.65$ regime has benefited tremendously from the explosive increase in the number of quasar spectra available for blind DLA surveys. The N12 statistical sample now includes 3408 DLAs from SDSS DR9. Of course, there is absolutely no hope for a similarly sized sample in the UV in the foreseeable future. Nevertheless, we can make progress by using the metal-line proxies for DLAs that we have developed. With over 40,000 \MgII\ systems now identified in SDSS spectra (Quider et al. 2011; Zhu \& Menard 2013; Seyffert et al. 2013; Sardane et al. 2017, in prep.), the statistics of \MgII\ systems are known to high statistical accuracy, with uncertainties most certainly dominated by systematics. As described in RTN06, if the fraction of DLAs in a well-defined \MgII\ sample is known, then the \MgII\ statistics can be used to determine the incidence of DLAs. Therefore, the UV surveys that we have carried out in the past (RTN06) and those that we describe below were carefully designed to select \MgII\ systems that have a high probability of being DLAs. Apart from enabling the statistical properties of DLAs to be determined, these also provide a valuable resource for follow-up studies of individual neutral gas reservoirs at low redshift. This includes their metallicities, dust contents, molecular fractions, star formation, kinematics, associated galaxies, and clustering as a function of redshift. 

\subsection{The HST ACS Prism Survey}

The details of the survey and measurements of absorption lines are presented in Turnshek et al. (2015). Here we briefly describe the survey and its results.  When STIS on HST failed in 2004 and COS had yet to be installed, we had the opportunity to make use of the low-resolution ACS-HRC-PR200L prism to perform a survey for DLAs in \MgII\ systems between redshifts $0.42<z<0.70$. This redshift interval was chosen because it was well-matched to the sensitivity of the PR200L. Despite the low resolution, $\sim 200$ at 1730 \AA\ and $\sim 100$ at 2070 \AA, it was still possible to detect and measure Ly$\alpha$ lines with DLA column densities ($N_{\rm HI} \ge 2 \times 10^{20}$ atoms cm$^{-2}$). Equally importantly, it was possible to determine which systems were not DLAs. We found 35 high probability DLAs, which were defined as those absorption lines with measured column densities within $1\sigma$ of $1 \times 10^{20}$ atoms cm$^{-2}$.   Formally, the sample included 26 DLAs with $N_{\rm HI} \ge 2 \times 10^{20}$ atoms cm$^{-2}$. For the 61 \MgII\ systems with \Wmi$ \ge 0.3$ \AA, the spectrum was of high enough signal-to-noise ratio to definitively conclude that there was no DLA line present at the \MgII\ redshift. For twelve others, the spectra could not be used to determine the nature of the Ly$\alpha$ line. Among these twelve, we include a \MgII\ system which could have a very large \HI\ column density, $N_{\rm HI}>2\times 10^{22}$ atoms cm$^{-2}$; we don't include this in our sample since its column density cannot be definitively determined from the prism data. Details of this sample are given in tables 1 and 2 from Turnshek et al. (2015). 

All the \MgII\ systems selected for this study had $W_0^{\lambda2796} \ge 1 $\AA. They were selected from an early version of the Pittsburgh SDSS Quasar \MgII\ Absorption-line Survey Catalog (Quider et al. 2011). A few sightlines had additional \MgII\ systems with $0.3$ \AA\ $< W_0^{\lambda2796} < 1$ \AA, and one system had $W_0^{\lambda2796} < 0.3$ \AA. As we explain in \S3, this distinction becomes important when assigning a selection criterion to a \MgII\ system in the calculation of the DLA incidence (see also RTN06).

\subsection{The GALEX Archival Survey}

The GALEX archive of quasar grism spectra is an additional resource that can be used to search for strong absorption features in the UV. 
Strong Ly$\alpha$ lines can easily be detected in the low-resolution grism data, which have $\lambda/\Delta\lambda \sim 200$ in the $1344-1786$ \AA\ FUV band  and $\sim 120$ in the $1771-2831$ \AA\ NUV band (Morrissey et al. 2007). We cross-matched our DR4, DR7, and DR9 SDSS \MgII\ catalogs (Quider et al. 2011; Sardane et al. 2017 in prep.), with the GALEX GR6 quasar spectroscopic catalog and found 122 matches of which 60 \MgII\ systems with $W_0^{\lambda2796} \ge 0.3$ \AA\ had useful spectra with signal-to-noise ratios greater than $\sim 3$ near the expected position of the Ly$\alpha$ line. We excluded GO programs in the GALEX archive which specifically targeted high-probability DLAs (e.g., Monier et al. 2009).  Of the 60 \MgII\ systems, 3 are DLAs. Two others were measured to be subDLAs. In the remaining spectra, it was only possible to definitively conclude that no strong absorption line was present at the expected position of Ly$\alpha$. These were not DLAs, but the possibility that they could be lower column density subDLAs could not be excluded. Table 1 gives details of the \MgII\ systems with UV spectra in the GALEX archive.

\begin{table*}
\caption{The GALEX Archival Sample\tablenotemark{a}}
\begin{tabular}{lccccccc}
\hline
\hline
Quasar & $z_{em}$ & $z_{abs}$ & \MgII\ $W^{\lambda2796}_{0}$  & \MgII\ $W^{\lambda2803}_{0}$ 
 & \MgI\ $W^{\lambda2852}_{0}$  & \FeII\ $W^{\lambda2600}_{0}$  & log ${\rm N_{\rm HI}}$\tablenotemark{b} \\
 &  & & (\AA)& (\AA) & (\AA)& (\AA)& atoms cm$^{-2}$ \\
\hline
J020114.3$-$090958	&	2.011	&	1.0023	&	1.504	$\pm$	0.071	&	1.366	$\pm$	0.074	&	0.218	$\pm$	0.067	&	0.772	$\pm$	0.073	&	\nodata	\\
J020329.8$-$091020	&	1.579	&	1.2166	&	0.435	$\pm$	0.041	&	0.361	$\pm$	0.045	&		$\le 0.1$				&	0.261	$\pm$	0.049	&	\nodata	\\
J020502.9$-$081020	&	1.293	&	0.9438	&	1.165	$\pm$	0.055	&	0.771	$\pm$	0.050	&		$\le 0.2$				&	0.365	$\pm$	0.049	&	\nodata	\\
J020502.9$-$081020	&	1.293	&	0.9612	&	0.429	$\pm$	0.050	&	0.201	$\pm$	0.051	&		$\le 0.2$				&			$\le 0.2$			&	\nodata	\\
J032605.6$-$064915	&	1.878	&	0.8842	&	0.421	$\pm$	0.074	&	0.515	$\pm$	0.068	&	0.027	$\pm$	0.062	&			$\le 0.2$			&	\nodata	\\
J074358.2+323512	&	0.906	&	0.7185	&	1.750	$\pm$	0.020	&	1.604	$\pm$	0.021	&	0.357	$\pm$	0.032	&	1.267	$\pm$	0.022	&	\nodata	\\
J074451.3+292005	&	1.183	&	1.0628	&	1.182	$\pm$	0.023	&	1.111	$\pm$	0.024	&	0.480	$\pm$	0.026	&	0.770	$\pm$	0.024	&	\nodata   \\
J083136.9+035414	&	1.136	&	0.824	&	0.777	$\pm$	0.109	&	0.741	$\pm$	0.099	&		$\le 0.3$				&	0.150	$\pm$	0.101	&	\nodata	\\
J083136.9+035414	&	1.136	&	1.0871	&	0.771	$\pm$	0.118	&	0.525	$\pm$	0.117	&	0.058	$\pm$	0.081	&	0.312	$\pm$	0.097	&	\nodata	\\
J084539.2+172311	&	1.295	&	0.9077	&	0.571	$\pm$	0.107	&	0.539	$\pm$	0.076	&	0.127	$\pm$	0.081	&	0.110	$\pm$	0.097	&	\nodata	\\
J100024.6+023149	&	1.311	&	0.7314	&	0.547	$\pm$	0.066	&	0.436	$\pm$	0.078	&	0.112	$\pm$	0.075	&	0.241	$\pm$	0.064	&	\nodata	\\
J100024.6+023149	&	1.311	&	0.9009	&	1.246	$\pm$	0.071	&	1.097	$\pm$	0.071	&	0.451	$\pm$	0.096	&	0.714	$\pm$	0.071	& 20.08$\pm$0.08\\
J104245.0+114638	&	1.080      &	1.0334	&	1.252	$\pm$	0.070	&	1.188	$\pm$	0.061	&	0.219	$\pm$	0.070	&	0.831	$\pm$	0.053	&	\nodata	\\
J104513.8$-$010246	&	0.971	&	0.6212	&	0.361	$\pm$	0.045	&	0.221	$\pm$	0.043	&	0.108	$\pm$	0.038	&	0.080	$\pm$	0.043	&	\nodata	\\
J104656.7+054150	&	1.234	&	0.7127	&	1.177	$\pm$	0.058	&	0.795	$\pm$	0.058	&	0.246	$\pm$	0.061	&	0.707	$\pm$	0.055	&	\nodata	\\
J104840.1+053551	&	1.973	&	0.9881	&	1.189	$\pm$	0.130	&	0.783	$\pm$	0.129	&		$\le 0.8$				&	1.039	$\pm$	0.15		&	\nodata	\\
J104932.2+050532	&	1.115	&	0.6626	&	1.143	$\pm$	0.148	&	0.919	$\pm$	0.205	&	0.210	$\pm$	0.153	&	0.599	$\pm$	0.246	&	\nodata	\\
J105158.5+590652	&	1.822	&	0.8174	&	1.614	$\pm$	0.091	&	1.177	$\pm$	0.088	&	0.260	$\pm$	0.080	&	0.822	$\pm$	0.094	&	\nodata	\\
J110211.8+284041	&	1.747	&	1.1516	&	0.847	$\pm$	0.104	&	0.698	$\pm$	0.116	&			$\le 0.3$			&	0.574	$\pm$	0.125	& 20.52$\pm$0.08	\\
J113221.9+651813	&	1.130	&	0.8333	&	0.521	$\pm$	0.065	&	0.633	$\pm$	0.076	&	0.249	$\pm$	0.060	&	0.277	$\pm$	0.070	&	\nodata	\\
J113221.9+651813	&	1.130	&	1.0092	&	1.078	$\pm$	0.076	&	0.965	$\pm$	0.076	&		\nodata				&	0.025	$\pm$	0.057	&	\nodata	\\
J121054.6+202202	&	0.979	&	0.9331	&	0.403	$\pm$	0.049	&	0.325	$\pm$	0.056	&	0.097	$\pm$	0.049	&			$\le 0.5$			&	\nodata	\\
J121404.1+330945	&	1.598	&	0.7546	&	0.400	$\pm$	0.052	&	0.185	$\pm$	0.048	&	0.145 	$\pm$ 	0.112 	&			$\le 0.2$			&	\nodata	\\
J122028.0+092826	&	1.082	&	0.8863	&	0.601	$\pm$	0.070	&	0.405	$\pm$	0.077	&		$\le 0.3$				&	0.166	$\pm$	0.071	&	\nodata	\\
J122031.7+094853	&	0.991	&	0.7683	&	0.543	$\pm$	0.087	&	0.296	$\pm$	0.072	&		$\le 0.2$				&			$\le 0.2$			&	\nodata	\\
J122424.4+300125	&	1.192	&	0.8246	&	1.101	$\pm$	0.142	&	0.753	$\pm$	0.150	&	0.313 	$\pm$ 	0.149 	&	0.184 	$\pm$ 	0.159 	&	\nodata	\\
J123116.4+120023	&	1.413	&	1.0960	&	0.401	$\pm$	0.040	&	0.305	$\pm$	0.041	&	0.049	$\pm$	0.037	&	0.260	$\pm$	0.037	&	\nodata	\\
J123508.6+254453	&	1.019	&	0.9668	&	0.381	$\pm$	0.065	&	0.438	$\pm$	0.065	&			$\le 0.3$			&			$\le 0.3$			&	\nodata	\\
J123934.1+623320	&	1.659	&	0.8832	&	1.704	$\pm$	0.144	&	1.240	$\pm$	0.117	&		$\le 0.45$			&	0.908	$\pm$	0.147	&	\nodata	\\
J124039.3+084251	&	1.098	&	1.0127	&	1.070	$\pm$	0.073	&	0.740	$\pm$	0.069	&	0.235	$\pm$	0.081	&	0.319	$\pm$	0.072	&	\nodata	\\
J124025.1+085818	&	1.269	&	0.8852	&	1.713	$\pm$	0.083	&	1.299	$\pm$	0.100	&	0.340	$\pm$	0.095	&	0.887	$\pm$	0.107	&	\nodata	\\
J124044.6+330348	&	0.812	&	0.6119	&	0.685	$\pm$	0.044	&	0.693	$\pm$	0.047	&	0.170 	$\pm$ 	0.053 	&	0.522	$\pm$ 	0.075	&	\nodata	\\
J124126.1+615804	&	1.162	&	1.1289	&	3.476	$\pm$	0.092	&	2.921	$\pm$	0.077	&	0.386	$\pm$	0.078	&	2.267	$\pm$	0.073	&	\nodata	\\
J131218.2+493316	&	1.620	&	0.8834	&	1.124	$\pm$	0.066	&	0.653	$\pm$	0.067	&		$\le 0.23$			&	0.212	$\pm$	0.061	&	\nodata	\\
J132330.5+293320	&	1.098	&	1.0449	&	2.758	$\pm$	0.129	&	2.560	$\pm$	0.122	&	0.661 	$\pm$	0.160	&	2.075	$\pm$	0.100	& 20.54$\pm$0.20	\\
J132613.7+285144	&	1.299	&	0.9423	&	0.419	$\pm$	0.053	&	0.295	$\pm$	0.053	&	0.171 	$\pm$ 	0.060	&	0.261 	$\pm$ 	0.049 	&	\nodata	\\
J134917.4+460311	&	1.704	&	1.0957	&	2.115	$\pm$	0.167	&	2.177	$\pm$	0.161	&	0.756	$\pm$	0.116	&	1.451	$\pm$	0.086	&	\nodata	\\
J140023.1+433852	&	1.127	&	0.7220	&	0.786	$\pm$	0.056	&	0.506	$\pm$	0.051	&	0.177	$\pm$	0.053	&	0.237	$\pm$	0.055	&	\nodata	\\
J140023.1+433852	&	1.127	&	0.9464	&	0.429	$\pm$	0.054	&	0.429	$\pm$	0.047	&		$\le 0.16$			&	0.189	$\pm$	0.047	&	\nodata	\\
J141549.7+005357	&	1.042	&	0.8676	&	0.373	$\pm$	0.033	&	0.280	$\pm$	0.033	&		$\le 0.1$				&			$\le 0.1$			&	\nodata	\\
J141551.1+012242	&	1.239	&	0.9612	&	0.556	$\pm$	0.056	&	0.351	$\pm$	0.050	&	0.042	$\pm$	0.048	&	0.131	$\pm$	0.044	&	\nodata	\\
J141838.4+522359	&	1.121	&	1.0232	&	1.483	$\pm$	0.121	&	0.844	$\pm$	0.113	&	0.138	$\pm$	0.137	&	0.451	$\pm$	0.088	&	\nodata	\\
J142315.3+375219	&	1.823	&	0.8353	&	1.922	$\pm$	0.062	&	1.686	$\pm$	0.068	&	0.653	$\pm$	0.072	&	1.400	$\pm$	0.068	&	\nodata	\\
J142501.4+382101	&	1.145	&	0.8212	&	0.488	$\pm$	0.029	&	0.413	$\pm$	0.031	&		$\le 0.1$				&	0.131	$\pm$	0.029	&	\nodata	\\
J143624.3+353709	&	0.768	&	0.3922	&	1.050	$\pm$	0.164	&	0.782	$\pm$	0.162	&		$\le 0.4$				&			\nodata			&	\nodata	\\
J143617.8+353726	&	1.448	&	0.8185	&	0.910	$\pm$	0.148	&	0.676	$\pm$	0.119	&		$\le 0.4$				&			$\le 0.3$			&	\nodata	\\
J144313.3+094910	&	1.192	&	0.9260	&	0.934	$\pm$	0.035	&	0.594	$\pm$	0.034	&	0.142	$\pm$	0.038	&	0.316	$\pm$	0.035	& 19.90$\pm$0.11\\
J153658.3+343149	&	0.890	&	0.8080	&	1.435	$\pm$	0.115	&	0.801	$\pm$	0.127	&	0.562	$\pm$	0.117	&	0.698	$\pm$	0.110	& 20.48$\pm$0.15	\\
J155251.3+314944	&	1.036	&	1.0023	&	1.513	$\pm$	0.027	&	1.256	$\pm$	0.029	&	0.202	$\pm$	0.044	&	0.745	$\pm$	0.032	&	\nodata	\\
J155454.8+110943	&	2.327	&	0.8402	&	0.652	$\pm$	0.064	&	0.430	$\pm$	0.054	&	0.126 	$\pm$ 	0.106 	&	0.310 	$\pm$ 	0.090 	&	\nodata	\\
J160714.1+492331	&	0.887	&	0.8410	&	0.373	$\pm$	0.037	&	0.288	$\pm$	0.038	&		\nodata				&	0.086	$\pm$	0.039	&	\nodata	\\
J160839.1+354226	&	1.042	&	0.7851	&	1.029	$\pm$	0.044	&	0.748	$\pm$	0.049	&	0.267	$\pm$	0.075	&	0.411	$\pm$	0.058	&	\nodata	\\
J160855.4+435259	&	0.933	&	0.7195	&	0.608	$\pm$	0.044	&	0.456	$\pm$	0.040	&	0.091	$\pm$	0.046	&	0.266	$\pm$	0.042	&	\nodata	\\
J160919.0+434727	&	1.211	&	1.1362	&	0.934	$\pm$	0.112	&	0.530	$\pm$	0.102	&	0.125	$\pm$	0.082	&	0.183	$\pm$	0.084	&	\nodata	\\
J161817.7+330900	&	1.147	&	0.8883	&	0.417	$\pm$	0.032	&	0.267	$\pm$	0.035	&	0.048	$\pm$	0.036	&	0.112	$\pm$	0.032	&	\nodata	\\
J161902.5+303051	&	1.288	&	1.0635	&	0.801	$\pm$	0.026	&	0.684	$\pm$	0.026	&	0.136	$\pm$	0.030	&	0.302	$\pm$	0.028	&	\nodata	\\
J171748.4+594820	&	0.763	&	0.3745	&	0.657	$\pm$	0.137	&	0.677	$\pm$	0.118	&	0.089	$\pm$	0.081	&			\nodata			&	\nodata	\\
J171748.4+594820	&	0.763	&	0.6078	&	0.332	$\pm$	0.067	&	0.209	$\pm$	0.054	&		$\le 0.2$				&			$\le 0.2$			&	\nodata	\\
J213042.3+003632	&	1.263	&	0.7771	&	1.569	$\pm$	0.105	&	1.192	$\pm$	0.130	&	0.135	$\pm$	0.111	&	0.537	$\pm$	0.097	&	\nodata	\\
J235731.7$-$101507	&	1.290	&	0.8809	&	1.542	$\pm$	0.055	&	1.556	$\pm$	0.055	&	0.288	$\pm$	0.067	&	1.083	$\pm$	0.062	&	\nodata\\
\hline
\end{tabular}
\vspace{-0.2in}
\tablenotetext{a}{Where indicated, upper limits are given at the 3$\sigma$ level.}
\tablenotetext{b}{Blank entries indicate that the Ly$\alpha$ line could not be measured, but is certainly not strong enough to be a DLA.}

\end{table*}

\subsection{The MMT-HST COS Survey}

Nestor et al. (2006) presented a low-redshift \MgII\ catalog that was obtained using the MMT blue spectrograph with the aim of extending the \MgII\ distribution to lower redshifts than accessible with the SDSS. They detected 140 \MgII\ systems with $W^{\lambda2796}_{0} \ge 0.1$ \AA\ at $z\ge 0.15$. A sample of $z_{abs}\le 0.4$ systems in UV-bright quasars was then selected for follow up with HST-COS in Cycle 19 (GO 12593, Nestor PI), and one of the aims of CoIs Rao and Turnshek was to characterize the neutral gas properties of the absorbers. Quasars with the possibility of Lyman limits in their spectra due to higher redshift absorption systems were eliminated, as were those with faint (FUV $> 21.5$) or unmeasured GALEX FUV fluxes. Three additional quasar spectra were available in the HST archives for use in the program. Measurements of the \MgII\ absorbers that can be included in our statistical sample are given in Table 2. 

We note that four additional \MgII\ absorbers that did not satisfy our selection criteria for finding \MgII-selected DLAs in a statistically unbiased way were also observed in GO 12593. One of them has $W^{\lambda2796}_{0} < 0.3$ \AA\ and is not a DLA. The other three were not from the MMT survey, but are very strong \MgII\ absorbers from our SDSS \MgII\ catalog with $W^{\lambda2796}_{0} \ge 2$ \AA; none of these three are part of our unbiased final sample since we never established a $W^{\lambda2796}_{0} \ge 2$ \AA\ cut in our DLA surveys. Two of the three were found to be DLAs. The $z_{abs} = 0.3955$, $W^{\lambda2796}_{0} = 2.32\pm 0.04$ \AA, system towards the quasar J123200.02-022404.7 (also known as PKS 1229$-$02) is a known 21 cm absorber, for which we measure $N_{\rm HI} = (5.0\pm 0.3) \times 10^{20}$ atoms cm$^{-2}$. The other DLA is the $z_{abs}=0.3969$, $W^{\lambda2796}_{0} = 2.23\pm 0.08$ \AA, system towards the quasar J125142.99$+$463734.7 with $N_{\rm HI} = (3.0\pm 0.2) \times 10^{20}$ atoms cm$^{-2}$. 

\begin{table*}
\caption{The $z_{abs} \le 0.4$ MMT-HST COS Sample}
\begin{tabular}{lccccccc}
\hline
\hline
Quasar & $z_{em}$ & $z_{abs}$ & \MgII\ $W^{\lambda2796}_{0}$  & \MgII\ $W^{\lambda2803}_{0}$ 
 & \MgI\ $W^{\lambda2852}_{0}$  & $\log {\rm N_{\rm HI}}$ \\
 &  & & (\AA)& (\AA) & (\AA)& atoms cm$^{-2}$ \\
\hline
J092554.70+400414.1 & 0.471 & 0.2471 & 0.966$\pm$0.044 & 0.850$\pm$0.043   & 0.289$\pm$0.051   & 19.55$\pm$0.15\tablenotemark{a}  \\
J094907.16+544510.5 & 1.369 & 0.3146 & 0.813$\pm$0.081 & 0.626$\pm$0.087   &0.222$\pm$0.117   &  19.08$\pm$0.09 \\
J094930.30$-$051454.0 & 1.098 & 0.1973 & 1.226$\pm$0.108 & 1.015$\pm$0.108   &0.143$\pm$0.093  & 19.36$\pm$0.09\\ 
J095000.73+483129.3 & 0.589 & 0.2117 & 0.578$\pm$0.063 & 0.398$\pm$0.063  & 0.096$\pm$0.057 & 16.18$\pm$0.06\tablenotemark{b}  \\
J095837.58+555053.1 & 1.021 & 0.2511 & 0.696$\pm$0.099 & 0.313$\pm$0.102  & 0.228$\pm$0.097  & 18.93$\pm$0.03\\
J100102.55+594414.3 & 0.752 & 0.3033 & 1.687$\pm$0.020 & 1.430$\pm$0.022  & 0.261$\pm$0.027  & 19.32$\pm$0.10\tablenotemark{a}  \\
J100736.06+363859.6 & 1.034 & 0.2106 & 0.473$\pm$0.056 & 0.307$\pm$0.057  & 0.096$\pm$0.065  &  18.85$\pm$0.06\\
J101703.49+592428.7 & 0.851 & 0.4039 & 0.675$\pm$0.052 & 0.554$\pm$0.041  & 0.083$\pm$0.039  &  18.90$\pm$0.08\\
J102117.47+343721.7 & 1.404 & 0.2611 & 2.158$\pm$0.045 & 1.786$\pm$0.049  &0.312$\pm$0.059  &  19.30$\pm$0.10\\
J102237.42+393150.5 & 0.605 & 0.2520 & 0.995$\pm$0.060 & 0.689$\pm$0.064  &0.123$\pm$0.068  &  18.95$\pm$0.10\\
J132325.25+343059.3 & 0.444 & 0.3529 & 0.889$\pm$0.048 & 0.666$\pm$0.038  & 0.279$\pm$0.040  &  19.23$\pm$0.08\\
J132815.61+524403.8 & 1.341 & 0.2683 & 0.300$\pm$0.046 & 0.193$\pm$0.044  &$\le 0.15\tablenotemark{c}$  & 18.40$\pm$0.20 \\
J140035.92+553534.4 & 0.835 & 0.3648 & 3.184$\pm$0.029 & 3.073$\pm$0.029  & 0.425$\pm$0.043  &  20.00$\pm$0.11\\
J141036.81+295550.9 & 0.570 & 0.3570 & 0.633$\pm$0.067 & 0.351$\pm$0.053  & 0.108$\pm$0.058  & 18.70$\pm$0.15 \\
J165931.93+373529.0 & 0.775 & 0.1994 & 0.842$\pm$0.066 & 0.795$\pm$0.078  & 0.276$\pm$0.069  &  19.30$\pm$0.15\\
J215647.46+224249.9 & 1.290 & 0.3648 & 0.912$\pm$0.012 & 0.740$\pm$0.013  &0.083$\pm$0.013  &  18.70$\pm$0.17\\
\hline
\vspace{-0.4in}
\tablenotetext{a}{From Battisti et al. (2012).}
\tablenotetext{b}{From Tumlinson et al. (2013).}
\tablenotetext{c}{3$\sigma$ upper limit}
\end{tabular}
\end{table*}

\section{Statistical Results}

\subsection{Defining the MgII-DLA sample}

As noted in RTN06, the \MgII\ systems selected for follow-up observations in the UV were chosen in a variety of ways that evolved over a number of years. Each selection method was unbiased, in that it conformed to specific but well understood criteria based on prior results. Our goal was to thoroughly quantify the probability of finding a DLA as a function of \Wmi\ while also finding sufficient numbers of them to determine their statistics and make follow-up observations.  The UV surveys that we have undertaken since then were described in \S2. Our sample of \MgII\ systems, which have been followed up in the UV for the purpose of determining whether the system is a DLA, now includes 369 systems with \Wmi$\ge 0.3$ \AA: 197 systems in the RTN06 sample, 96 systems in the ACS-Prism sample (Turnshek et al. 2015), 60 systems in the GALEX Archival sample (Table 1), and 16 systems in the MMT-HST sample (Table 2). This search has yielded 70 DLAs, which is the largest unbiased sample of low-redshift DLAs that has been assembled for cosmological studies, but which still pales in comparison to the statistical sample of 3408 DLAs (out of over 6800 DLAs in total) in the high redshift sample of N12.

We use the same methodology to estimate the satistical properties of DLAs as described in RTN06, which we briefly summarize here. The initial searches for UV DLAs were in strong \MgII\ systems with $W_0^{\lambda2796} \ge 0.3$ \AA; this matched the \MgII\ survey threshold of Steidel \& Sargent (1992). We adopted this because, with the statistical incidence of \MgII\ systems known from the Steidel \& Sargent (1992) survey, we could use the fraction of DLAs in a well-defined \MgII\ sample to bootstrap from the \MgII\ incidence to obtain the DLA incidence. At the time, it was also known that all high-redshift, optically identified DLAs, had \MgII\ absorption that met this threshold. 
Based on the results from our archival survey, our earliest non-archival HST surveys, which involved obtaining new HST observations and described in RTN06, had \MgII\ thresholds of $W_0^{\lambda2796} \ge 0.6$ \AA.
We also used the strength of \FeII\ to further define subsamples where the DLA fraction would be high. Refining our sample selection in this manner ensured that we used HST time with the maximum efficiency possible in searching for the rare, low-redshift DLAs. Then, with the HST ACS-Prism survey, we selected systems with $W_0^{\lambda2796} \ge 1.0$ \AA, but no \FeII\ selection criterion. Since the number of GALEX spectra available in the archives is small, we did not place further constraints on the \MgII\ systems in the GALEX sample, but simply searched for DLAs in $W_0^{\lambda2796} \ge 0.3$ \AA\ systems. The MMT-HST survey sample also had a $W_0^{\lambda2796} \ge 0.3$ \AA\ threshold. 

We define these subsamples as follows:

 \begin{itemize}
\item[1.]{ \Wmi$\ge0.3$ \AA; }
\item[2.]{ \Wmi$\ge0.6$ \AA;}
\item[3.]{ \Wmi$\ge0.6$ \AA\ and \Wf$\ge 0.5$ \AA;}
\item[4.]{ \Wmi$\ge1.0$\AA\ and  \Wf$\ge0.5$ \AA; and }
\item[5.]{\Wmi$\ge1.0$\AA}
\end{itemize}

Subsample 1 includes  all systems surveyed in our initial archival survey (Rao \& Turnshek 2000), the GALEX archival sample, and the MMT-HST COS sample.  It also includes any additional systems that happened to fall along  quasar sightlines that were targeted due to the presence of another stronger system from subsamples 2, 3, 4, or 5. Subsamples 2 and 3 were mainly  targeted for observation in HST-Cycle 9, and subsample 4 includes systems found in SDSS-EDR spectra and observed in HST-Cycle 11. A few systems from the  SDSS-EDR sample have strong MgII and FeII, but have \Wmi$\lesssim 1.0$\AA;  these belong in subsample 3. The ACS-Prism sample belongs in subsample 5. The reader is referred to Turnshek et al. (2015) for the \MgII\ systems in this sample; we do not reproduce it here. The reason for carefully defining these subsamples is that the calculation of DLA incidence is based on the \MgII\ incidence, and the incidence of \MgII\ is rest equivalent width-dependent, which is considered in \S3.2.

\begin{figure*} 
\includegraphics[width=1.\columnwidth]{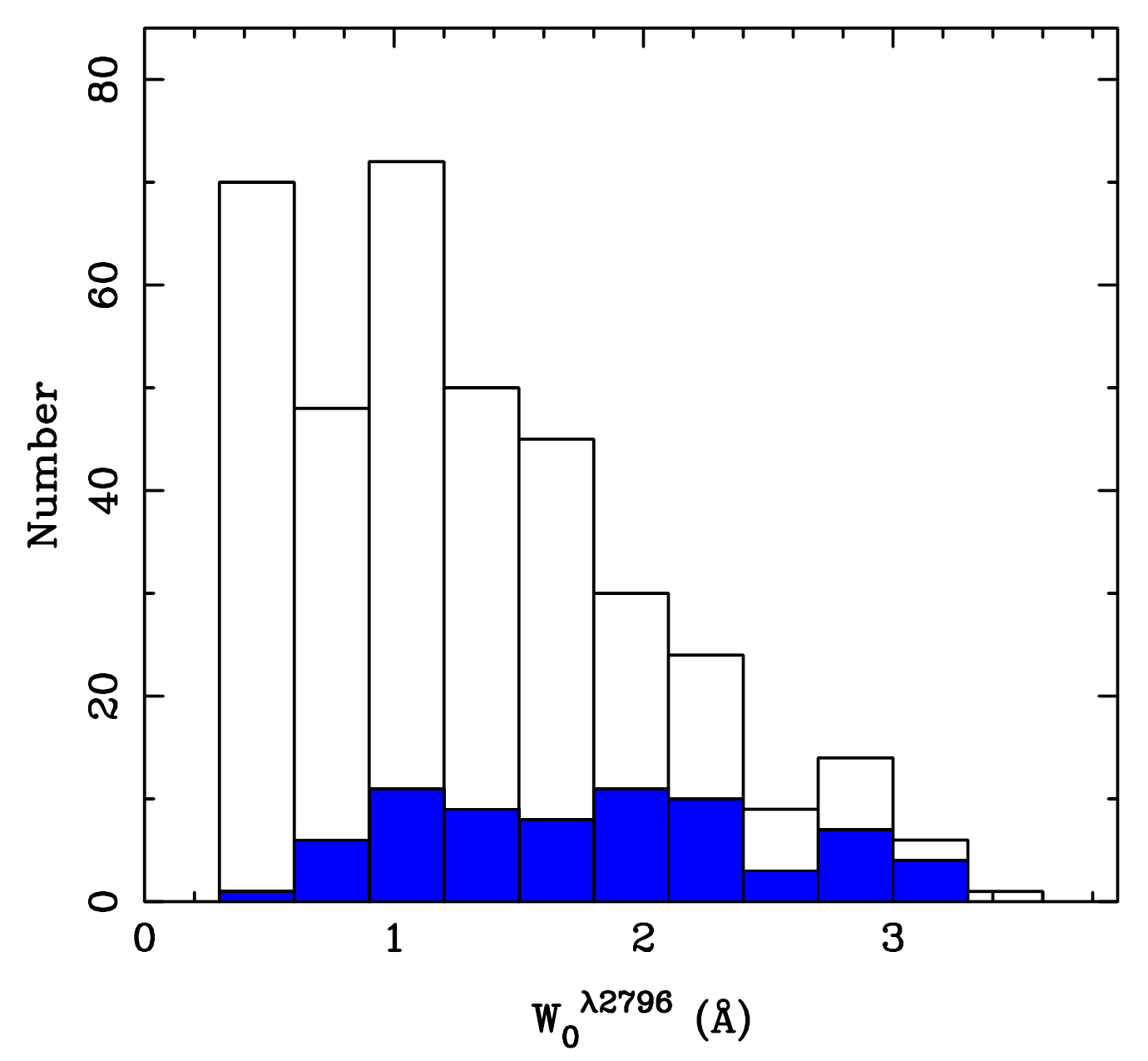}
\caption{Histogram showing the distribution of \MgII\ rest equivalent widths in our sample. The DLAs are the solid blue histogram.}
\label{Whist} 
\end{figure*}

\begin{figure*}
\includegraphics[width=1.\columnwidth]{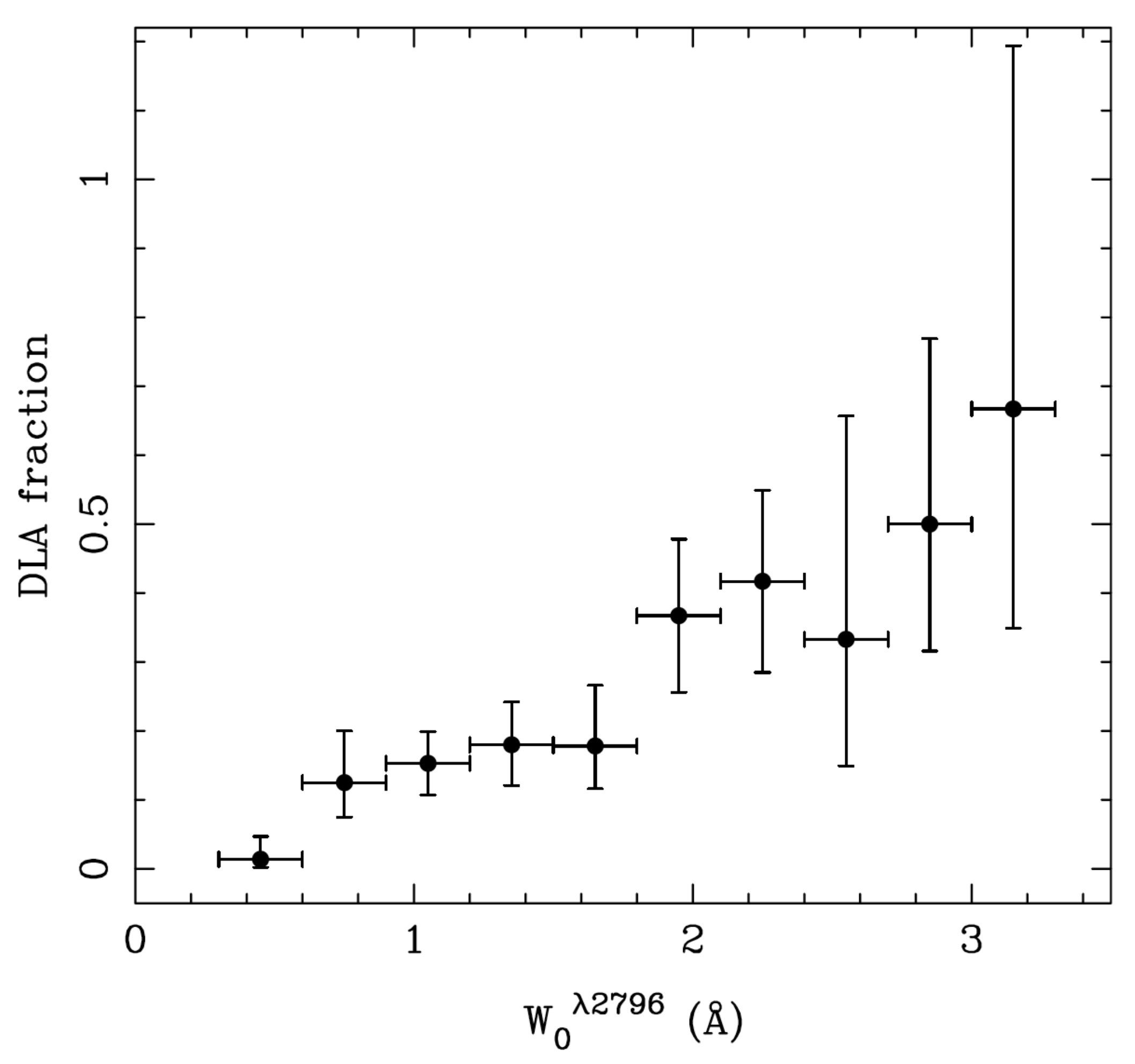}
\caption{The fraction of \MgII\ systems that are DLAs, which can also be interpreted as the probability of a \MgII\ system being a DLA, as a function of \Wmi. }
\label{DLAfraction}
\end{figure*}

Figure \ref{Whist} shows the distribution of \MgII\ rest equivalent widths in our sample. The DLAs are shaded in blue. We note that the lowest rest equivalent width bin, 0.3 \AA\ $\le$ \Wmi $< 0.6$ \AA, now has one DLA among 70 \MgII\ systems, whereas the RTN06 sample had none. The fraction of DLAs as a function of \Wmi\ increases from 1.4\% in the lowest \Wmi bin to 67\% in the bin centred at \Wmi$=3.15$ \AA. See Figure \ref{DLAfraction}, where Poisson errors with 68\% confidence limits are plotted. The last bin from Figure \ref{Whist} with 3.3 \AA\ $\le $ \Wmi $<3.6$ \AA\ is not plotted in Figure \ref{DLAfraction}. It includes one \MgII\ system which is not a DLA.

\subsection{The cosmic incidence of neutral gas, $n_{\rm{DLA}}$}

 The following is adapted from RTN06. The redshift number density of DLAs, $n_{\rm{DLA}}(z)$,  can be determined using the equation
\begin{equation}
n_{\rm DLA}(z) = \eta(z)\, n_{\rm MgII}(z),
\end{equation}
where $\eta(z)$ is the fraction of DLAs in a \MgII\ sample as a function of  redshift and $n_{\rm MgII}(z)$ is the redshift number density of \MgII\ systems.  Since our \MgII\ sample was assembled under various selection criteria  (see \S3.1), $n_{\rm MgII}(z)$ needs to be evaluated carefully. We can express $n_{\rm MgII}(z)$ for our sample as
\begin{equation}
n_{\rm MgII}(z)=\frac{1}{369}\sum_i w_i\, n_{{\rm MgII}_i}(z),
\end{equation}
where the sum is over all 369 systems, $w_i$ is a weighting factor that depends on the $i^{th}$ system's selection criterion for being included in the survey, and $n_{{\rm MgII}_i}(z)$ is the  $i^{th}$ system's  $dn/dz$ value calculated using the parametrization derived in NTR05 and updated in Sardane et al. (2017, in prep.):
\begin{equation}
dn/dz=N^*\,(1+z)^\alpha\,e^{-\frac{W_0}{W^*}(1+z)^{-\beta}},
\end{equation}
where  $W_0\equiv$ \Wmi\ is defined for simplicity and $N^*$, $W^*$, $\alpha$, and $\beta$  are constants.  In this expression, $dn/dz$ is the integrated redshift number density over all $W_0$ greater than the threshold $W_0$. For our calculation, the threshold $W_0$ is different for each of the five subsamples that comprise our total sample (see \S3.1 and RTN06). Thus, for example, a system that belongs  to subsample 1 has a threshold of $W_0 = 0.3$ \AA\ in Equation 3 and a weight of $w_i=1$ in Equation 2, while a system in subsample  2 has a threshold of  $W_0=0.6$ \AA\ and $w_i=1$, and a system in subsample 5 has a threshold of $W_0=1.0$ \AA\ and $w_i=1$. This follows since subsamples 1, 2, and 5 are purely \MgII\ selected without regard to the strength or presence of the \FeII\ $\lambda 2600$ line. On the other hand, a system that  belongs to subsample 3 has a threshold of $W_0=0.6$ \AA\ and $w_i=0.54$. This is because a \FeII\ $\lambda 2600$ criterion was used to select systems in addition to using a $W_0$ threshold, and 54\% of the 1,130  systems with $W_0 \ge 0.6$ \AA\ in the \MgII\ survey of NTR05 have \Wf $\ge 0.5$  \AA. Similarly, for systems in subsample 4 we have a threshold of $W_0=1.0$ \AA\ and $w_i=0.72$ because 72\% of the 781 systems with $W_0 \ge  1.0$ \AA\ in the \MgII\ survey of NTR05 have  \Wf$\ge 0.5$ \AA. For subsamples 3 and 4  we have assumed that  the fraction of \MgII\ systems that are also strong \FeII\ systems is independent of redshift. 

We have determined the \MgII\ parametrization for our updated SDSS DR4+DR7 \MgII\ catalog (Sardane et al. 2017, in prep.), and the values for the constants in Equation 3 are as follows:
$N^* = 1.015 \pm 0.006$, $W^* = 0.442 \pm 0.007$, $\alpha = 0.044 \pm 0.011$, and $\beta = 0.618 \pm 0.011$. These are consistent with the NTR05 values to within the errors. For redshifts $0.1<z<0.36$, our MMT survey for \MgII\ systems (Nestor 2004; Nestor et al. 2006) showed that the parametrization of the evolution of $W_0$ is consistent with that found for the $z>0.36$ SDSS \MgII\ systems with $W_0 \ge 0.3$ \AA\ and $W_0 \ge 0.6$ \AA. We therefore use Equation 3 with the above parameters for the entire redshift range of our sample ($0.11<z<1.65$.)

\begin{figure*}
\includegraphics[width=1.5\columnwidth]{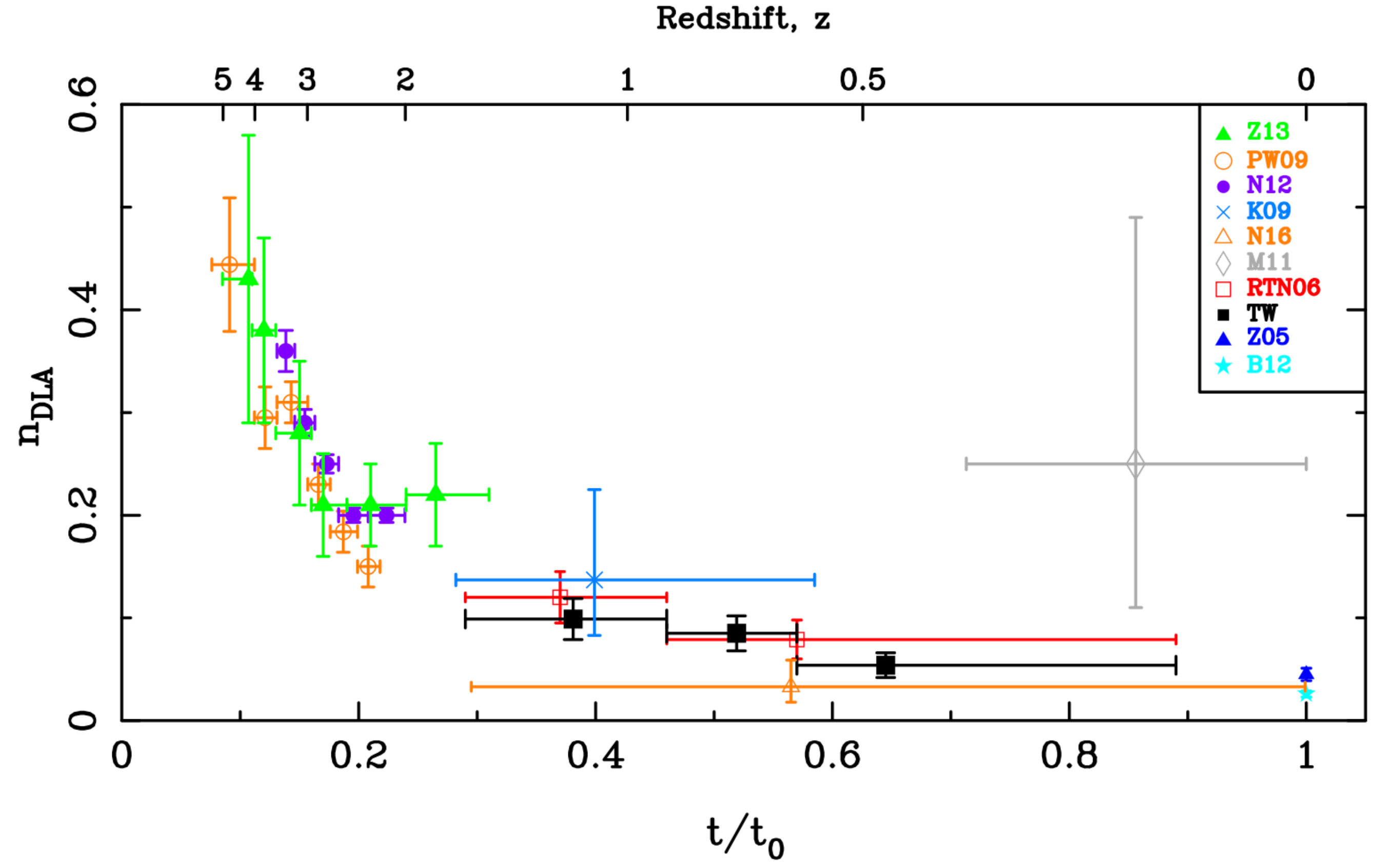}
\caption{Redshift number density of DLAs, $n_{\rm DLA}(z)$ plotted as a function of cosmic time with $t=t_0$ being the present epoch. Redshifts are noted along the top axis. A few key results from the literature are also plotted. References in the legend are as follows: Z13: Zafar et al. (2013); PW09: Prochaska \& Wolfe (2009);  N12: Noterdaeme et al. (2012); K09: Kanekar et al. (2009); N16: Neeleman et al. (2016); M11: Meiring et al. (2011); RTN06: Rao, Turnshek, \& Nestor (2006); TW: this work;  Z05: Zwaan et al. (2005); B12: Braun (2012).}
\label{dndtLegend}
\end{figure*}

\begin{figure*}
\includegraphics[width=1.5\columnwidth]{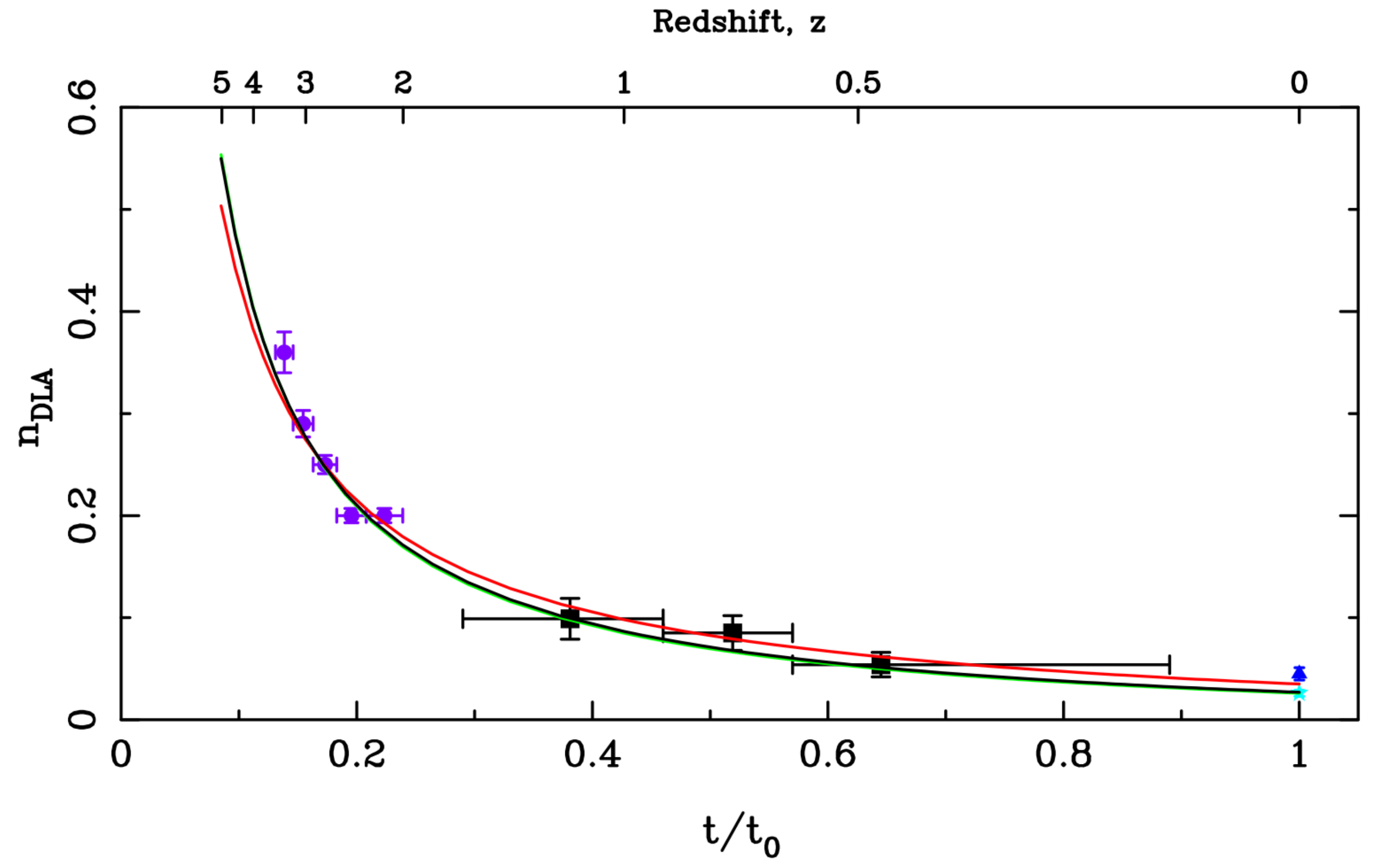}
\caption{Same as Figure \ref{dndtLegend}, but only the lowest uncertainty studies in the high- and low-redshift regimes are retained.  The $z>2$ data are from Noterdaeme et al. (2012) and the black squares are this work. The $z=0$ data points are  from Z05 (dark blue triangle) and  B12 (light blue star). The three curves are power law fits to the data of the form described in the text. The red curve includes the $z>0$ data points and the Z05 data point, the green curve includes the $z>0$ data points and the B12 data point, and the black curve includes only the $z>0$ data points. Note that the black and green curves very nearly overlap.}
\label{dndtfits}
\end{figure*}

Using the formalism described above, we find $n_{\rm DLA}(z)$ values as given in Table \ref{dndztable} and plotted as solid black squares in Figures \ref{dndtLegend} and \ref{dndtfits}. The redshift intervals were selected to include approximately equal numbers of \MgII\ systems in each bin. Figure \ref{dndtLegend} also shows data from the literature as described in the legend. We chose to plot $n_{\rm DLA}(z)$ as a function of time to highlight the large cosmic time interval probed by UV DLA surveys. While the general trend of these results indicates a steep decline in the cross-section of DLA absorbers from redshifts 5 to 2, there is less agreement among the various studies at low redshift because of small sample sizes. For example, Meiring et al. (2011) found 3 serendipitous DLAs in the COS-Halos survey (the grey diamond in Figure \ref{dndtLegend}); Neeleman et al. (2016) identified 4 DLAs in a blind survey of 463 quasars in the HST archives (the orange triangle). The blue cross is a 21 cm follow-up study of \MgII\ systems by Kanekar et al. (2009) in which they found 9 detections of 21 cm absorption among 55 \MgII\ systems. The red open squares are our previous results which included 41 DLAs in 197 \MgII\ systems (RTN06). The solid black squares are results from our current sample and include 70 DLAs in 369 \MgII\ systems. These new data confirm that the decline in DLA absorber cross-section continues to occur between redshifts $z=2$ to $z=0$. 

In Figure \ref{dndtfits} we only plot data with the smallest uncertainties in each redshift regime. At high redshifts these are the SDSS DR9 results from N12, who correct for incompleteness and false identifications in previous studies. The recent results of S\'anchez-Ram\'irez et al. (2016) for redshifts $1.6<z<5$ are consistent with the N12 results, albeit with larger uncertainties. The $z=0$ data points are from Z05, who used {\it Westerbork Synthesis Radio Telescope} \HI\ maps of a large sample of nearby galaxies to estimate $n_{\rm DLA}(z)$, and Braun (2012, henceforth B12) who used extremely high-resolution maps of Local Group galaxies and the local galaxy \HI\ mass function to calculate the \HI\ cross section as a function of $N_{\rm HI}$ after accounting for 21 cm beam size and opacity effects. The discrepancy between these two local estimates using 21 cm emission is almost a factor of two: $0.045\pm 0.006$ versus $0.026\pm 0.003$ from Z05 and B12, respectively. With the uncertainties in the $z>0$ data having improved considerably, we can now use these data to help inform and possibly constrain the $z\sim0$ values of $n_{\rm DLA}$. We show three power-law fits to the data in Figure \ref{dndtfits}. The red curve is a fit to the eight $z>0$ data points and the Z05 data point (dark blue triangle), the green curve is a fit that includes the $z>0$ data points and the B12 data point (light blue star), and the black curve is a fit to the $z>0$ data points alone. The fits are of the form 
\begin{equation}
n_{\rm DLA}(z) = n_0 (1+z)^\gamma,
\end{equation}
where $n_0 = 0.035\pm 0.007$ and $\gamma = 1.488 \pm 0.151$ for the $z>0$ $+$ Z05 data points,
$n_0=0.026\pm 0.004$ and $\gamma=1.707 \pm 0.108$ for the $z>0$ $+$ B12 data points,
and $n_0=0.027 \pm 0.007$ and $\gamma = 1.682 \pm 0.200$ for the $z>0$ data points alone. 

Extrapolation of the $z>0$ fit down to $z=0$ clearly favors the B12 determination of $n_{\rm DLA}$. All three fits are consistent with the $z>3.5$ data points shown in  Figure \ref{dndtLegend}. The $z>0$ power law is what we report for the form of the evolution of the incidence of neutral gas DLAs as a function of redshift for $0<z<5$, i.e.,
\begin{equation}
n_{\rm DLA}(z) = (0.027 \pm 0.007) (1+z)^{(1.682 \pm 0.200)}.
\end{equation}

\begin{table*}
\caption{UV-DLA statistics: $n_{\rm DLA}(z)$ and $\Omega_{\rm DLA}(z)$}
\begin{tabular}{ccccc}
\hline
\hline
Redshift & Mean $z$ & $n_{\rm DLA}(z)$ & $\left<N_{\rm HI}\right>$\tablenotemark{a} & $\Omega_{\rm DLA}(z)$\\
Interval & & & (cm$^{-2}$) & \\
\hline
0.11 - 0.61 & 0.464 & $0.054 \pm 0.012$ & $(1.32\pm 0.34)\times 10^{21}$ & $(7.7\pm2.6)\times 10^{-4}$\\
0.61 - 0.89 & 0.731 & $0.085 \pm 0.017$ & $(8.75\pm 1.34)\times 10^{20}$ & $(6.7\pm1.6)\times 10^{-4}$\\
0.89 - 1.65 & 1.172 & $0.099 \pm 0.020$ & $(1.02\pm 0.22)\times 10^{21}$ & $(7.5\pm2.9)\times 10^{-4}$\\
\hline
\vspace{-0.4in}
\tablenotetext{a}{Bootstrap errors are reported.}
\end{tabular}
\label{dndztable}
\end{table*}

\subsection{The cosmic mass density of neutral gas, $\Omega_{\rm DLA}$}

\begin{figure*}
\includegraphics[width=1.5\columnwidth]{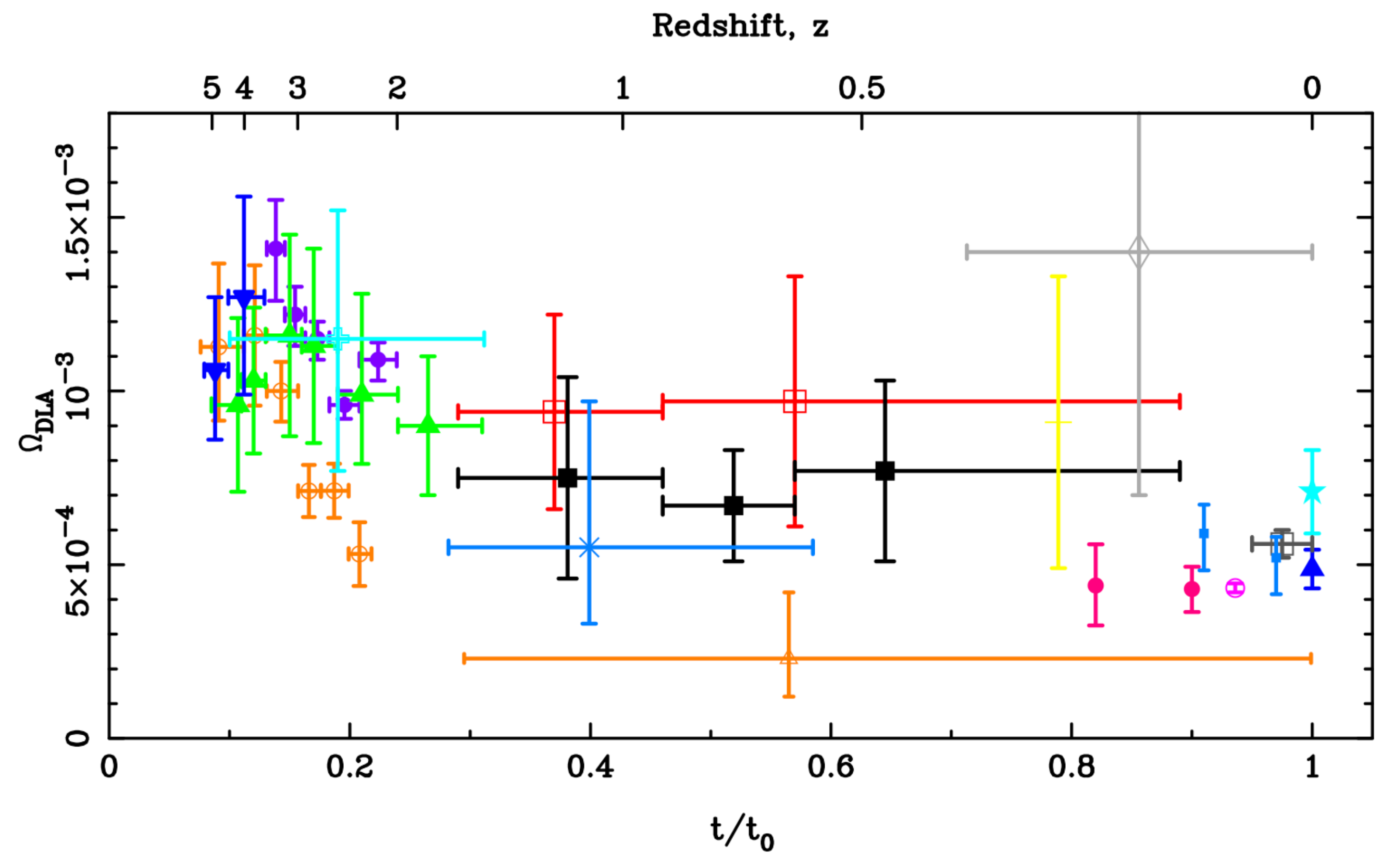}
\caption{The cosmological mass density of neutral gas as measured by DLAs, $\Omega_{\rm DLA}$, plotted as a function of cosmic time with $t=t_0$ being the present epoch. Redshifts are noted along the top axis. Many notable results from the literature are also plotted. Those in common with Figure \ref{dndtLegend} are plotted with the same symbols and colors. Our current results are solid black squares. Many references have been added. (These generally do not quote values for $n_{\rm DLA}(z)$.) From high to low redshift, these are Crighton et al. (2015, dark blue inverted triangles), Jorgenson et al. (2006, light blue open plus), Lah et al. (2007, yellow horizontal bar), Rhee et al. (2013, magenta filled circles), Hoppmann et al. (2015, purple open circle), Delhaize et al. (2013, small solid blue squares),  and Martin et al. (2010, dark grey open square). The large scatter among these various determinations highlights the difficulty in accurately measuring $\Omega_{\rm DLA}$.}
\label{Omt_all}
\end{figure*}

We compute $\Omega_{\rm DLA}$ as before (RTN06). DLA column density measurements, $N_{\rm HI}$, and the incidence of DLAs, $n_{\rm DLA}(z)$, can be used to determine $\Omega_{\rm DLA}(z)$ via the expression
\begin{equation}
\Omega_{\rm DLA}(z)= \frac{\mu m_{\rm H}}{\rho_c} \frac{H_0}{c} n_{\rm DLA}(z)
\left<N_{\rm HI}\right>  \frac{E(z)}{(1+z)^2},
\end{equation}
where
\begin{equation}
E(z)=(\Omega_M(1+z)^3 + (1-\Omega_M-\Omega_\Lambda)(1+z)^2 +
\Omega_\Lambda)^{1/2}.
\end{equation}
Again, the ``737'' cosmology is used to calculate $\Omega_{\rm DLA}$.  Also, $\mu=1.3$ corrects for a neutral gas composition of 75\% H and 25\% He by mass, $m_{\rm H}$ is the mass of the hydrogen atom,  $\rho_c$ is the critical mass density of the universe, and $\left<N_{\rm HI}\right>$ is the mean $N_{\rm HI}$ of DLAs in each redshift bin. The mean $N_{\rm HI}$ and $\Omega_{\rm DLA}$ values calculated in the same three redshift bins as $n_{\rm DLA}(z)$ are given in Table \ref{dndztable}.  Figure \ref{Omt_all} is a  plot of our current results, shown as solid black squares, as well as various results from the literature. The references are given in the captions to Figures  \ref{dndtLegend} and \ref{Omt_all}. The large scatter in the various determinations of $\Omega_{\rm DLA}$ is remarkable and highlights the inherent difficulty in estimating $\Omega_{\rm DLA}$, which is primarily due to small sample sizes and the fact that the mean \HI\ column density of a sample is dominated by the highest (and rarest) column density systems.  

A comparison of our previous results on $\Omega_{\rm DLA}$ (RTN06), shown as open red squares in Figure \ref{Omt_all}, with our current results shows that $\Omega_{\rm DLA}$ in the UV regime is now somewhat smaller, although the two determinations are consistent within the errors. As we show in \S3.4, this is due to the fact that our previous sample had a higher fraction of high column density systems, which skewed the column density distribution function higher for $\log N_{\rm HI} > 21.5$, but leaving it within $2\sigma$ of the high-redshift data. Now, with a sample that's nearly twice as large, we no longer see evidence of this. This result illustrates the fact that the value of $\Omega_{\rm DLA}$ is dominated by the highest column densities in the sample, and minor changes in the column density distribution at the high $N_{\rm HI}$ end can affect $\Omega_{\rm DLA}$. Thus, it is probably worth noting that our previous value of $\Omega_{\rm DLA}$ (RTN06) was never ``biased high'' by our \MgII-selection methods. Instead it should be recognized that small number statistics at the high $N_{\rm HI}$ end lead to significant uncertainties; these uncertainties at low redshift can be reduced by measuring larger samples of \MgII-selected DLAs. 

\begin{figure*}
\includegraphics[width=1.5\columnwidth]{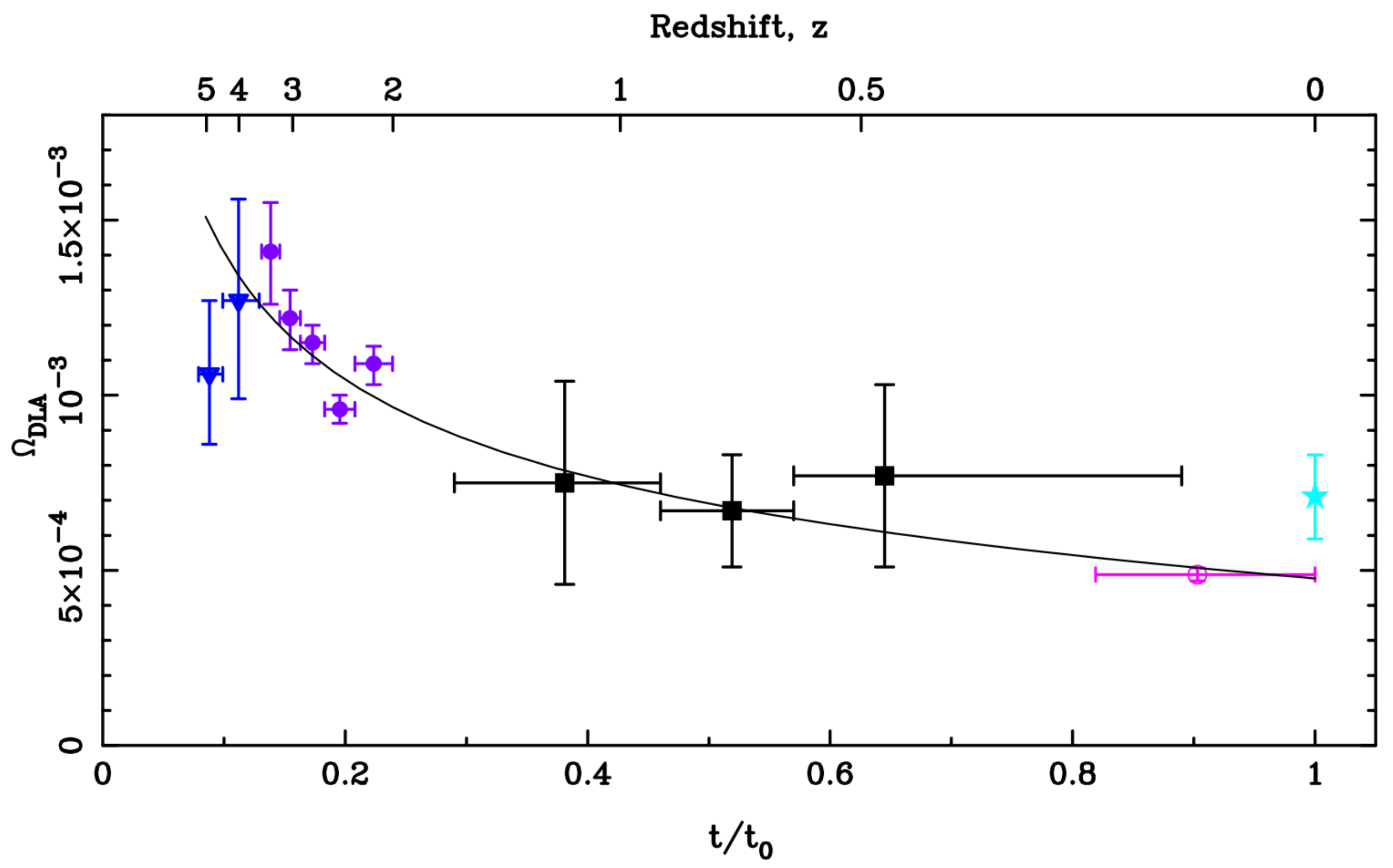}
\caption{Same as Figure \ref{Omt_all}, but with only the following data plotted: Crighton et al. (2015), N12, this work, Hoppman et al. (2015), and B12. See text. The symbols and colors are the same as in Figure \ref{Omt_all}. The fit only includes the $2<z<3.5$ data points of N12 and our  current results for $0.11<z<1.65$, and is of the form shown in Equation 8.}
\label{Omtfit}
\end{figure*}

In Figure \ref{Omtfit} we plot only one set of results for each redshift range: Crighton et al. (2015) for $z\ge 4$, N12 for $2<z<3.5$, our current results for $0.11<z<1.65$, the Hoppmann et al. (2015) data point that is a combined estimate for all 21 cm emission results for $0<z<0.2$ but which does not include the Braun (2012) result, and the Braun (2012) data point. 

As we did for $n_{\rm DLA}(z)$, we fit a power law to only the N12 data points and our current results. We derive
\begin{equation}
\Omega_{\rm DLA}(z) = (4.77 \pm 1.60)\times10^{-4} (1 + z)^{(0.64\pm 0.27)},
\end{equation}
which is shown as the solid curve in Figure \ref{Omtfit}. We extrapolate this curve to $z=0$, but it is clear that the uncertainties in the low-redshift data points are such that no conclusion should be drawn as to how the $z>0$ DLA results may inform or constrain the 21 cm emission results at $z=0$, unlike extrapolation of the $n_{\rm DLA}(z)$ fit to $z=0$ (Figure \ref{dndtfits}.)

\subsection{The DLA column density distribution, $f(N)$}

\begin{figure*}
\includegraphics[width=1.5\columnwidth]{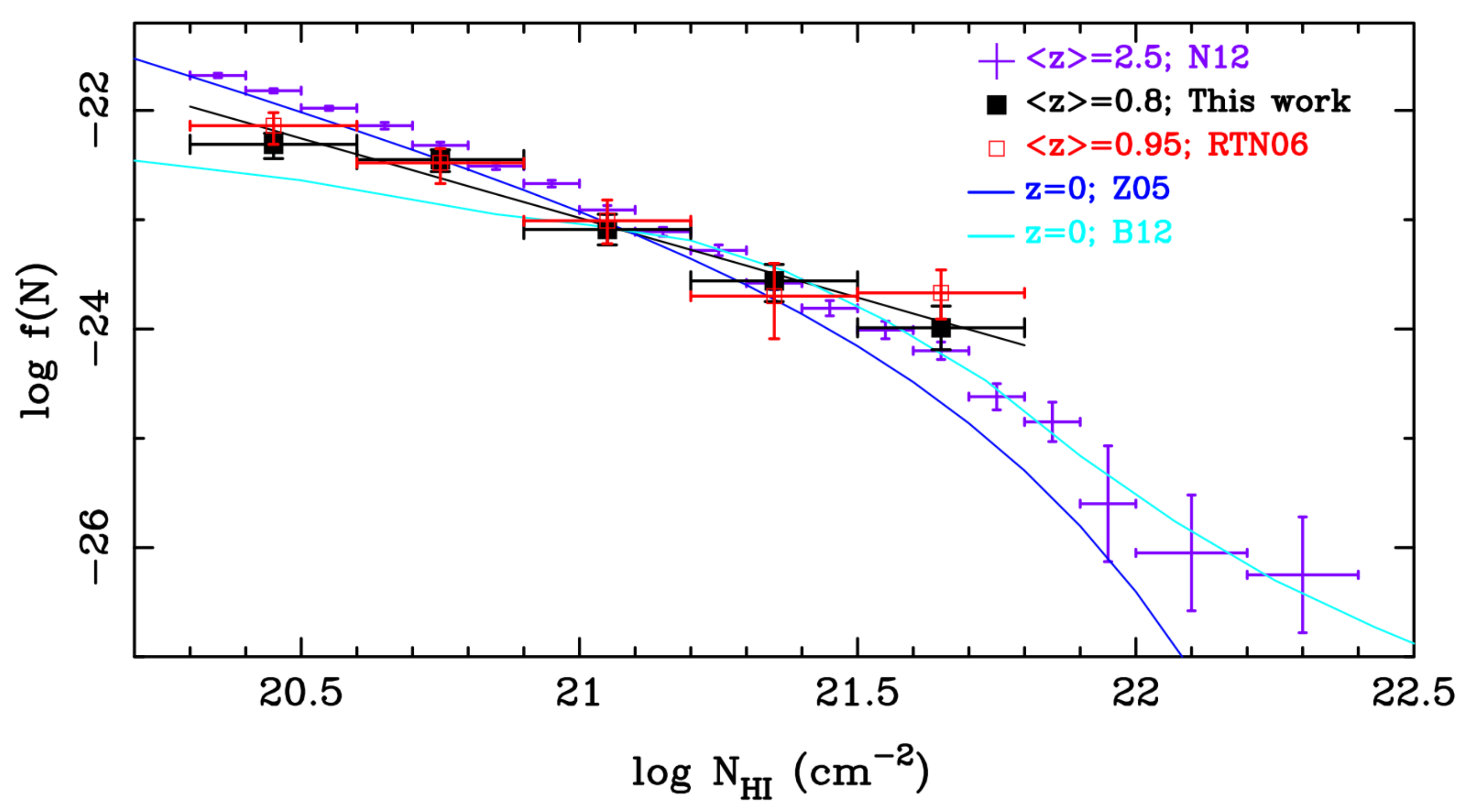}
\caption{The column density distribution of DLAs. Our current results are shown as black squares, with the best-fitting power law shown as the black solid line. The fit has a slope of $-1.46 \pm 0.20$. Our earlier RTN06 results are the open red squares; the excess of high $N_{\rm HI}$ systems is not seen in the larger updated sample. The low $N_{\rm HI}$ tail of the distribution is seen to decrease as one progresses from high redshift (N12), to low redshift (this study), to $z=0$ (B12), although to establish this we would need to more clearly understand why the results of Z05 and B12 disagree. At the highest $N_{\rm HI}$ values our results are more consistent with both the $z=0$ 21 cm emission results of B12 as well as the high redshift results of N12. }
\label{fNplot}
\end{figure*}

\begin{table*}
\caption{$f(N)$ at $\left < z\right > = 0.8$}
\begin{tabular}{cc}
\hline
\hline
$\log~N_{\rm HI}$ range & $\log f(N)$ \\
\hline
20.3 $-$ 20.6 & $-22.31^{+0.10}_{-0.13}$\\
20.6 $-$ 20.9 & $-22.45^{+0.09}_{-0.11}$\\
20.9 $-$ 21.2 & $-23.09 \pm 0.14$\\
21.2 $-$ 21.5 & $-23.56^{+0.15}_{-0.19}$\\
21.5 $-$ 21.8 & $-23.99\pm 0.20 $\\
\hline
\end{tabular}
\end{table*}

Figure \ref{fNplot} is a plot of the column density distribution of DLAs. The black solid squares were calculated using the expression
\begin{equation}
f(N,z) = n_{\rm DLA}(z) \frac{E(z)}{(1+z)^2} \frac{y(N,z)}{\Delta N},
\end{equation}
where $y(N,z)$ is the fraction of DLAs with column densities between $N$ and $N+\Delta N$ at redshift $z$, and $E(z)$ is given by Equation 7, as was the case in RTN06.  The $f(N)$ values, listed in Table 4,  were calculated at the mean redshift of our total sample, $\left < z \right > = 0.789$, and using the redshift number density of DLAs for all systems in the redshift interval $0.11<z<1.65$, $n_{\rm DLA}(z) = 0.078 \pm 0.009$. The black solid line in Figure \ref{fNplot} is a fit to the data points and has a slope of $-1.46 \pm 0.20$. A comparison with our previous results from RTN06 shows only minor differences, with the largest deviation being in the last $\log N_{\rm HI}$ bin. The higher fraction of high column density systems that we noted in RTN06 has now disappeared.
The current $f(N)$ value in our highest $N_{\rm HI}$ bin is consistent with both the high-redshift value of N12 and the $z=0$ value of B12. (Other more recent estimates of $f(N)$ at high redshift are consistent with the N12 results, but have larger error bars due to smaller samples, and we do not plot them here.) We have not detected any systems with $\log N_{\rm HI} > 21.8$ cm$^{-2}$, and therefore cannot comment on the high-$N_{\rm HI}$ tail of the $f(N)$ distribution at low redshift beyond this value. Nonetheless, the high $N_{\rm HI}$ end of the $f(N)$ distribution appears not to have evolved with redshift, while the low $N_{\rm HI}$ end, on the other hand, is significantly different in all three redshift regimes if the B12 data are used at $z=0$. If the Z05 result is used instead, then the $f(N)$ distributions diverge at both ends: the low-redshift data points from the other two at low values of $N_{\rm HI}$ and the $z=0$ curve from the other two at high values of $N_{\rm HI}$. The interpretation of these trends requires better data, or better understanding of the data, surprisingly, at $z=0$. The B12 work highlights the need for caution when interpreting 21 cm maps of galaxies in terms of quasar absorption line information, and this caution applies both at high and low \HI\ column densities. At the high $N_{\rm HI}$ end, since the conversion of 21 cm brightness temperature assumes that the gas is in the optically thin regime, the true value of $N_{\rm HI}$ may be higher than estimated from emission maps due to saturation effects. At the low column density end, higher resolution 21 cm maps  ($\sim$100 pc for the B12 data versus 1.4 kpc for the Z05 data), which better approximate quasar pencil beam surveys, show a decrease in the relative numbers of low-$N_{\rm HI}$ sightlines. This is likely due to the presence of higher column density gas within larger beams. However, since the B12 results were estimated based on a scaling of the properties of three Local Group galaxies to the \HI\ mass function of the local universe, there may be systematic uncertainties in the overall normalization of the derived cosmological parameters.

\section{On the \MgII\ selection method}

One often-noted concern about using the \MgII-selection method to conduct a UV survey for DLAs is that it may preferentially select systems with higher metallicities, and that we might be missing lower metallicity DLAs which would show up as systems with \Wmi $< 0.3$ \AA, or perhaps they might have no detectable \MgII\ absorption. Another concern has been that by using strong \MgII\ as a proxy for DLAs, our sample could include a higher fraction of high $N_{\rm HI}$ systems, thus causing our derived $\Omega_{\rm DLA}$ to be biased high.  We address these concerns below in reverse order. In short, there is absolutely no evidence for these effects at $z<1.65$.

\subsection{Is  $\left < N_{\rm HI} \right>$ biased high?}

In RTN06 we presented the \Wmi\ and $N_{\rm HI}$ distributions of our sample of 41 DLAs in 197 \MgII\ systems, where figures 3 and 4 of RTN06 showed that the fraction of \MgII\ systems that are DLAs increases with increasing \Wmi. This result is also true for our current expanded sample (Figure \ref{DLAfraction}). The data points in figure 4 of RTN06 showed that the mean $N_{\rm HI}$ value, $\left < N_{\rm HI} \right>$, for all the \MgII\ systems (DLAs and non-DLAs) remains constant as a function of \Wmi\ for \Wmi$ \ge 0 .6$ \AA. We did not find any DLAs at 0.3 \AA\ $\le$ \Wmi\ $< 0.6$ \AA\footnote{We also reiterate that no DLA has been found with \Wmi $< 0.3$ \AA, and thus, bins with lower values of \Wmi\ do not contribute to the cosmic neutral-gas density of the universe.}.  Unfortunately we cannot make a similar plot using {\it all} of the new \MgII\ absorbers in our current sample because we do not have measured $N_{\rm HI}$ values for most of the new subDLAs. In figure 5 of RTN06 we showed $\left < N_{\rm HI} \right>$ for only DLAs in that sample; it hinted at a {\it decreasing} trend of $\left < N_{\rm HI} \right>$ with increasing \Wmi, if at all. In fact, the error bars, and the low point in the bin at \Wmi $\sim 1.35$ \AA, showed the mean $N_{\rm HI}$ to be consistent with a value of $\sim1\times 10^{21}$ cm$^{-2}$ as a function of \Wmi. There was no evidence in the prior data that bins at larger \Wmi\ had higher $\left < N_{\rm HI} \right>$. The results did show, of course, a higher probability of larger \Wmi\ systems hosting a DLA, but the $\left < N_{\rm HI} \right>$ of such DLAs was clearly not higher than the  $\left < N_{\rm HI} \right>$ of lower \Wmi\ DLAs. 

\begin{figure*}
\includegraphics[width=1.5\columnwidth]{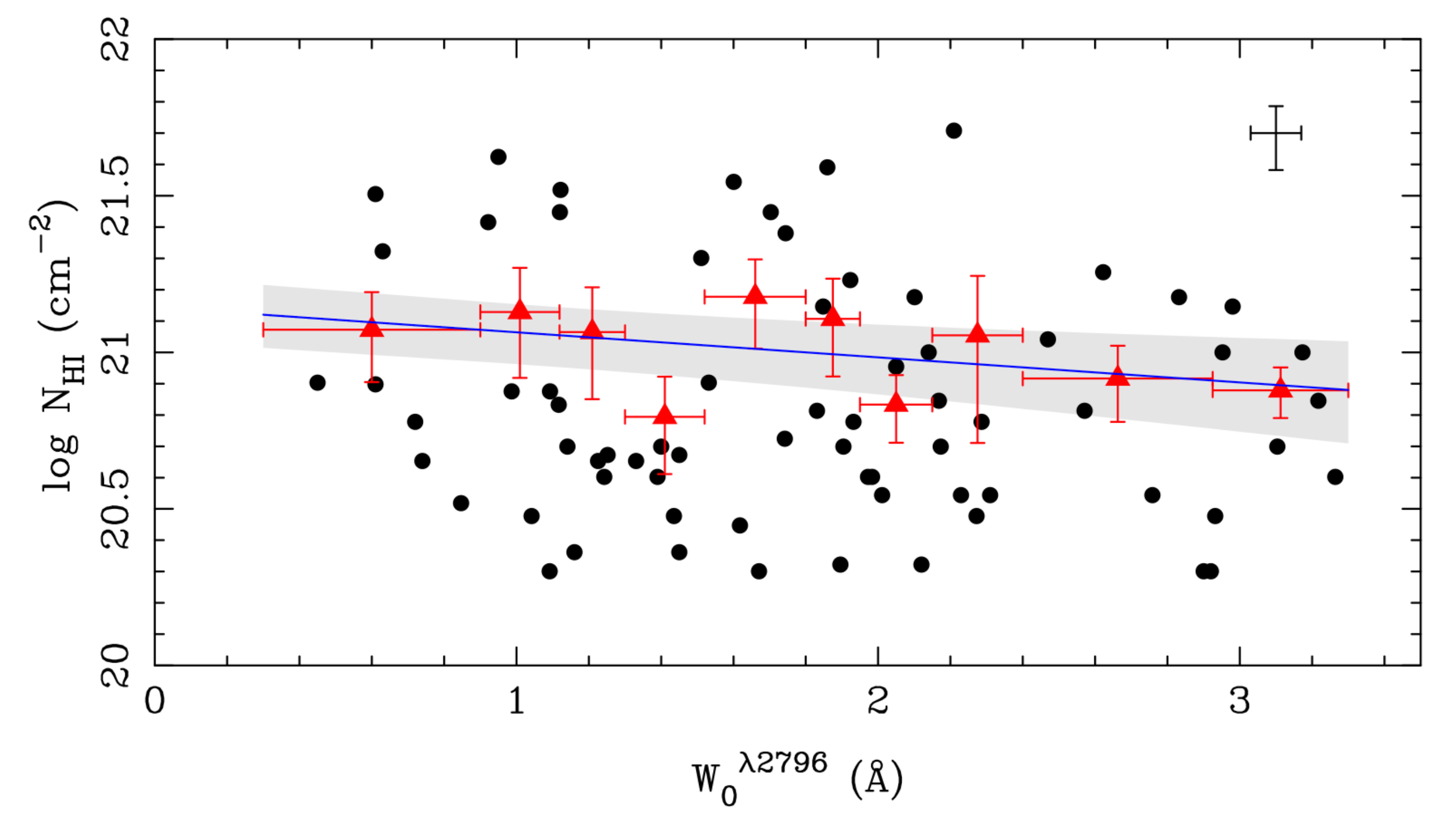}
\caption{Log $N_{\rm HI}$ versus \Wmi\ for the 70 DLAs in our sample. The average error bars of the data points are shown in the top-right corner. Mean \HI\ column densities are shown as red points for 10 bins chosen to have an equal number of DLAs in each bin. Bootstrap errors are shown. The solid line is a fit to these points and has a slope of $-0.08 \pm 0.05$. The gray region represents the 95\% confidence band for the fit. }
\label{NHIvsWDLAs}
\end{figure*}

The above results remain true for our new, updated sample. Figure \ref{NHIvsWDLAs} shows the plot of $N_{\rm HI}$ versus \Wmi\ for our current sample of 70 DLAs. Once again, there is no evidence for an increasing trend of $N_{\rm HI}$ with increasing \Wmi. Also, we now have one system in the updated sample with 0.3 \AA\ $\le$ \Wmi\ $< 0.6$ \AA, and this system has $N_{\rm HI} = (8 \pm 2) \times 10^{20}$ cm$^{-2}$. The red points in Figure \ref{NHIvsWDLAs} are mean values of $N_{\rm HI}$ calculated in 10 bins of 7 DLAs each. Bootstrap errors are shown. The solid line is a fit to these data points that has a slope of $-0.08 \pm 0.05$, consistent with no correlation between $\left <N_{\rm HI} \right >$ and \Wmi. The grey region represents the 95\% confidence band for the fit. We see that the lack of very high $N_{\rm HI}$ systems in the top right corner of Figure \ref{NHIvsWDLAs} persists, and is driving the marginally negative slope, although this is likely to be due to small number statistics. 

We should clarify several points. Our latter surveys for DLAs with HST preferentially selected \MgII\ systems with  \Wmi\ $\ge 0.6$ \AA\ since results from the initial survey indicated that this would lead to a higher probability of finding a DLA.  Then, for example, had there been a positive correlation between \Wmi\ and $N_{\rm HI}$, this type of selection could have resulted in preferentially finding DLAs with higher values of $N_{\rm HI}$, i.e., when calculating $\left < N_{\rm HI} \right>$ for the entire sample, $\left < N_{\rm HI} \right>$ would have been biased high. Thus, even though $n_{\rm DLA}$ would not have been affected (i.e., biased) because it is calculated using the relative incidence of \MgII\ systems (Equations 2 and 3), the calculation of $\Omega_{\rm DLA}$ using Equation 6 would have been biased high had we used a value for $\left < N_{\rm HI} \right>$ that was biased high. To correct for this we would then have needed to calculate $\Omega_{\rm DLA}$ in parts as a function of \Wmi. However, as we showed above, $\left < N_{\rm HI}\right>$ is independent of \Wmi\ for all values of \Wmi\ where DLAs are detected, and therefore such a ``bias'' does not exist in our previous sample nor in our new, updated \MgII-selected sample over the redshift interval of $0.11 < z < 1.65$ (the UV DLA regime). If anything, if the minimal negative correlation shown in Figure \ref{NHIvsWDLAs} is real, it would mean a negative bias at the highest \Wmi\ values, which would have required a slight (but insignificant) upward correction for $\Omega_{\rm DLA}$.

\subsection{Is there a metallicity bias?}

Since our selection is based on a `metal,' \MgII, the \HI\ gas that is being traced by \MgII\ systems is not `pristine.' There must have been an epoch when DLAs were not enriched enough to have \MgII\ absorption, in which case, our selection method would miss them. There may also be pockets of high-density neutral gas that have very low metallicity at redshifts $z<1.65$ and which do not have \MgII\ above the threshold of our survey. These would also be missed. However, there is no evidence for this. All DLAs observed thus far at all redshifts have \MgII\ or other low ionization absorption (e.g., Turnshek et al. 1989, Lu et al. 1993, Lu \& Wolfe 1994, Wolfe et al. 2005). In addition, the high redshift ($2<z<6$) MgII-DLA study of Matejek et al. (2013) has shown that all DLAs in their sample have \Wmi$>0.4$ \AA. Thus, metal-enrichment existed at the highest redshifts of the DLAs found in their study ($z\sim 5.3$). With its large oscillator strength, the \MgII$\lambda2796$ line becomes saturated at relatively low column densities, and thus, its rest equivalent width cannot be used as a measure of metallicity.  The DLAs found by Matejek et al. (2013) were measured to have metallicities, e.g., [Si/H], with lower limits down to a few thousandths solar, with several lower limits measured at a tenth solar. Thus, there are no known DLAs that are extremely metal poor and that do not show strong \MgII\ absorption even at the highest redshifts studied thus far. Matejek et al. (2013) also found that the metallicities of these MgII-DLAs are not different than those of the general population of DLAs at those redshifts, i.e., \MgII\ selection does not bias the DLA sample towards higher metallicities. They conclude that the observed metallicities are not inconsistent with the hypothesis that the two groups trace the same population of absorbers. 

Until recently, almost all of the DLA and sub-DLA metallicity measurements at low redshift were for systems identified in our surveys. These measurements showed that there is a steady increase in the metallicity of the universe from high redshift to the present epoch, and that subDLAs generally have higher metallicities than DLAs (e.g., Kulkarni et al. 2007, 2010, Som et al. 2015, Quiret et al. 2016 and references therein). If there was a metallicity bias in the selection of DLAs and subDLAs at low redshift, then the observed increase in metallicity with time would have to be corrected for this bias. Dessauges-Zavadsky et al. (2009) speculated that the high metallicities of subDLAs at low redshift were likely due to a bias introduced by our \MgII\ selection of DLAs and subDLAs. The DLAs in the sample of Nestor et al. (2008) also had higher mean metallicities than non-MgII selected DLAs by about 0.1 dex. However, the Nestor et al. (2008) sample mainly included systems with \Wmi$>1$ \AA. 

\begin{figure*}
\includegraphics[width=1.5\columnwidth]{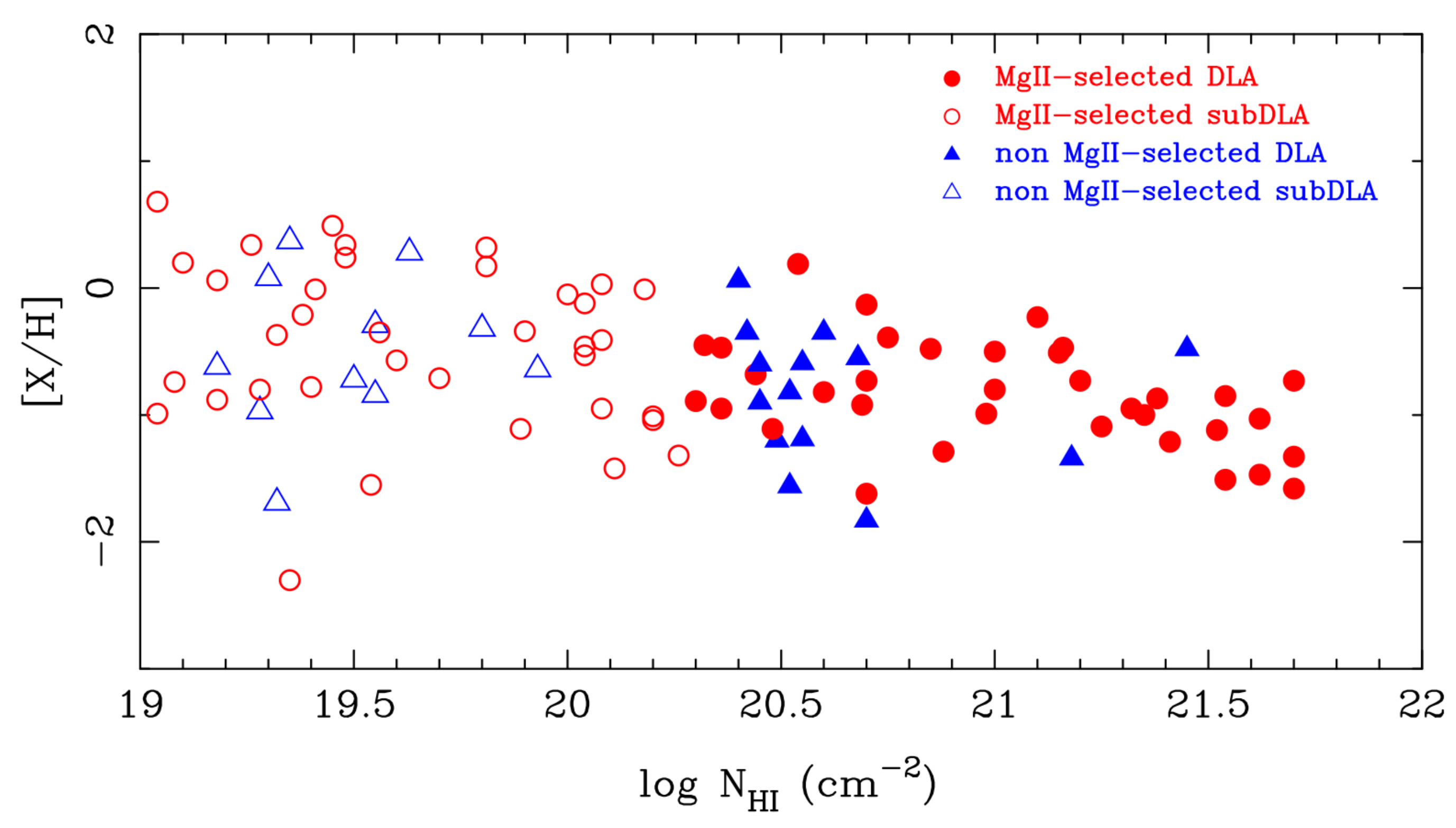}
\caption{ Metallicities of DLAs and subDLAs at $z<1.65$ versus log \HI\ column density. Metallicity measurements are from a variety of literature sources. The majority are from Quiret et al. (2016) and references therein.  Solid and open red circles are \MgII-selected DLAs and subDLAs, respectively. These include systems from our sample as well as from R. Becker's \MgII-selected HST DLA survey (GO 9051; Rao et al. 2005; Lacy et al. 2003; Meiring et al. 2006) and a 21 cm-DLA survey of \MgII\ systems from Ellison et al. (2012). Blue solid and open triangles are non-\MgII-selected DLAs and subDLAs.}
\label{ZvsNHI}
\end{figure*}

\begin{figure*}
\includegraphics[width=1.5\columnwidth]{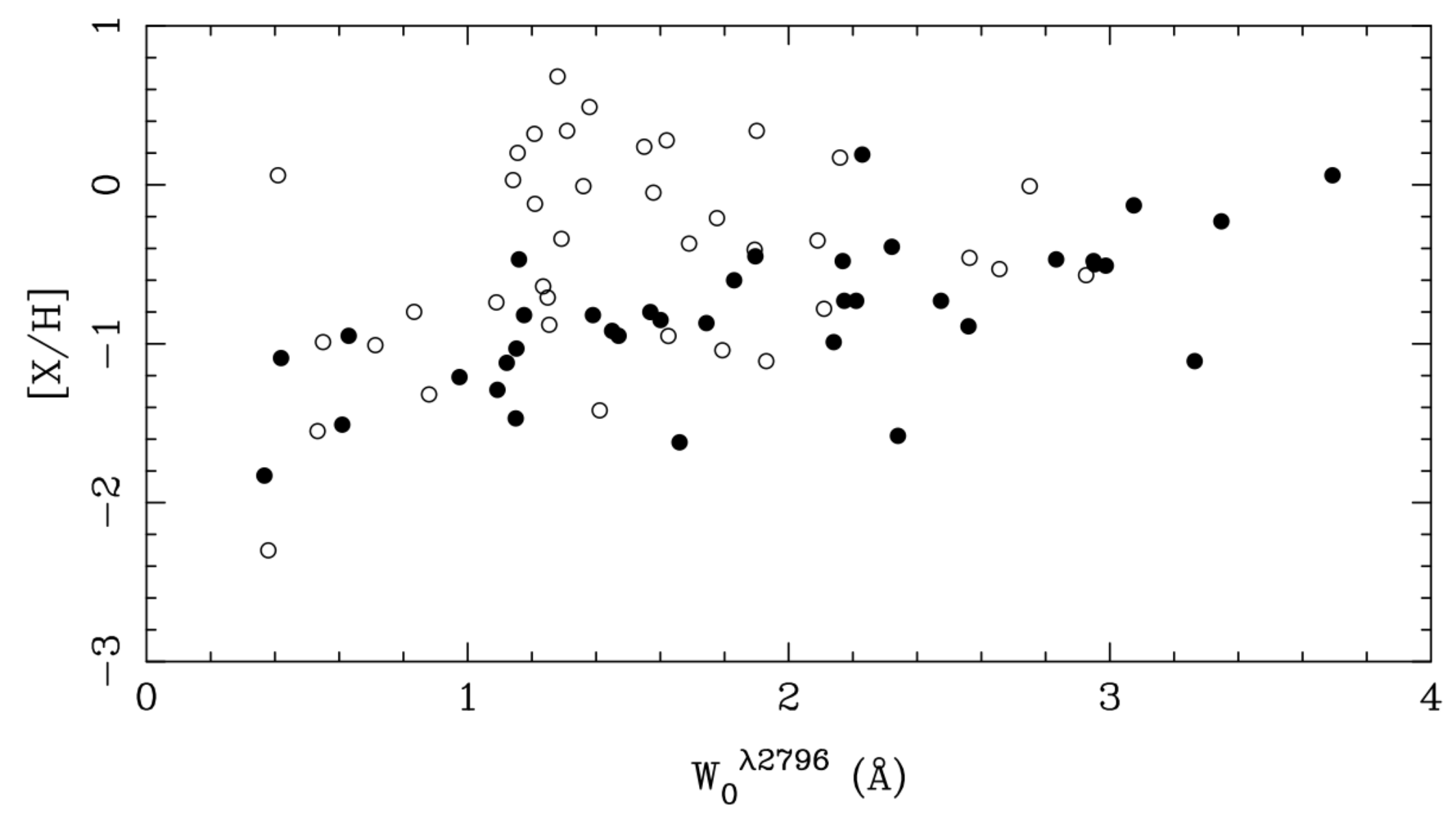}
\caption{Metallicities of DLAs (solid circles) and subDLAs (open circles) at $z<1.65$ versus \MgII$\lambda2796$ rest equivalent width. The well-known increase in metallicity with \Wmi\ is evident in this sample, with the DLAs showing a tighter correlation. All known DLAs have \Wmi$\ge 0.3$ \AA, and all $z<1.65$ DLAs have metallicities [X/H]$>-2$. Extraplation of the DLA metallicity-\Wmi\ correlation to 
\Wmi$ = 0.3$ \AA\ is consistent with the existence of a floor in DLA \Wmi\ and [X/H] values.}
\label{XvsW}
\end{figure*}

\begin{figure*}
\includegraphics[width=1.5\columnwidth]{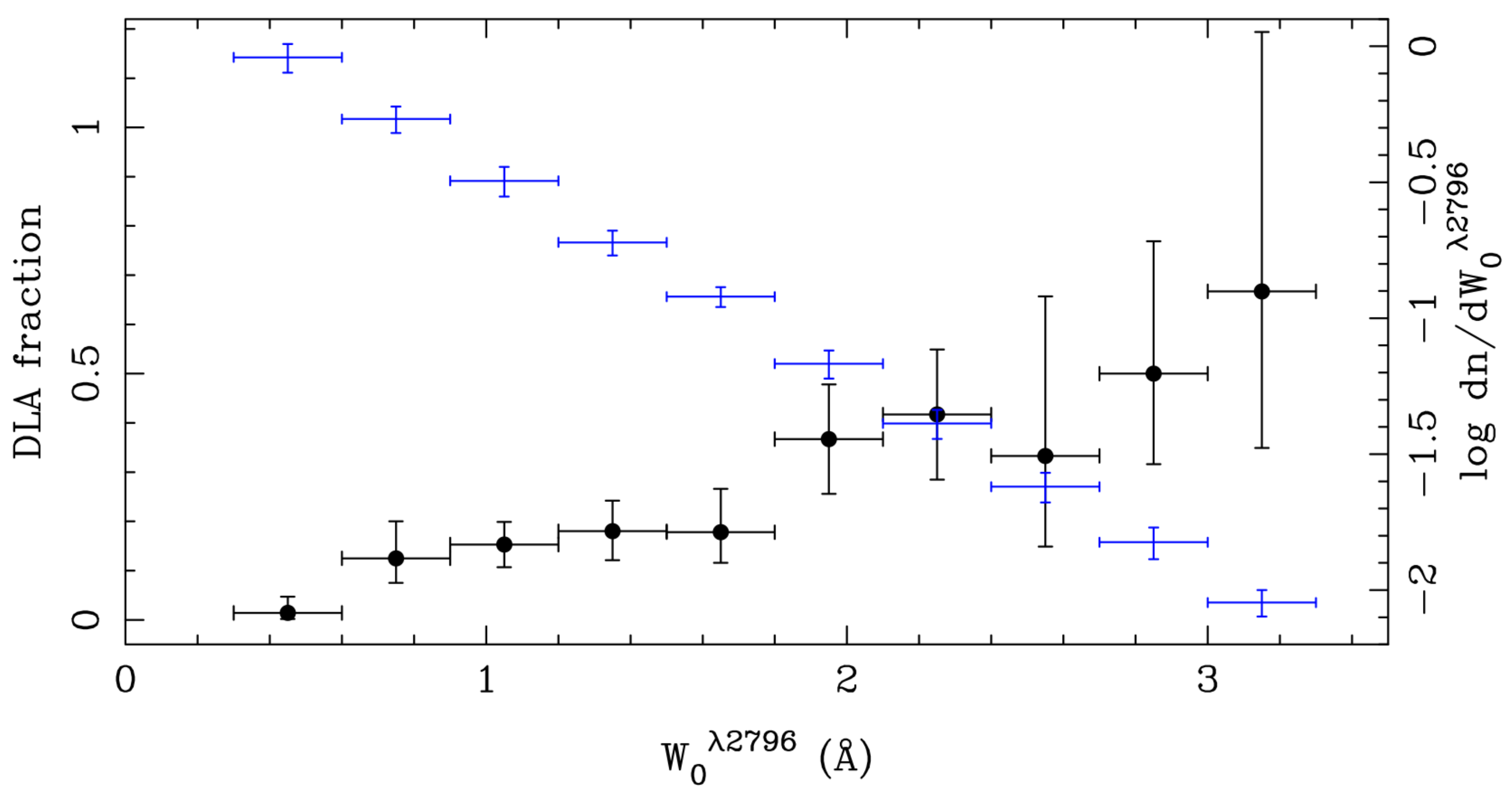}
\caption{The black data points are the DLA fraction in \MgII\ systems as a function of \Wmi\, and are the same as in Figure \ref{DLAfraction}. The blue points and the right vertical axis represent the cosmic \Wmi\ distribution of \MgII\ systems for \Wmi$\ge 0.3$ \AA\ for the mean redshift of our sample, $\left< z\right >=0.8$, from NTR05. }
\label{DLAfracdndW}
\end{figure*}

\begin{figure*}
\includegraphics[width=1.5\columnwidth]{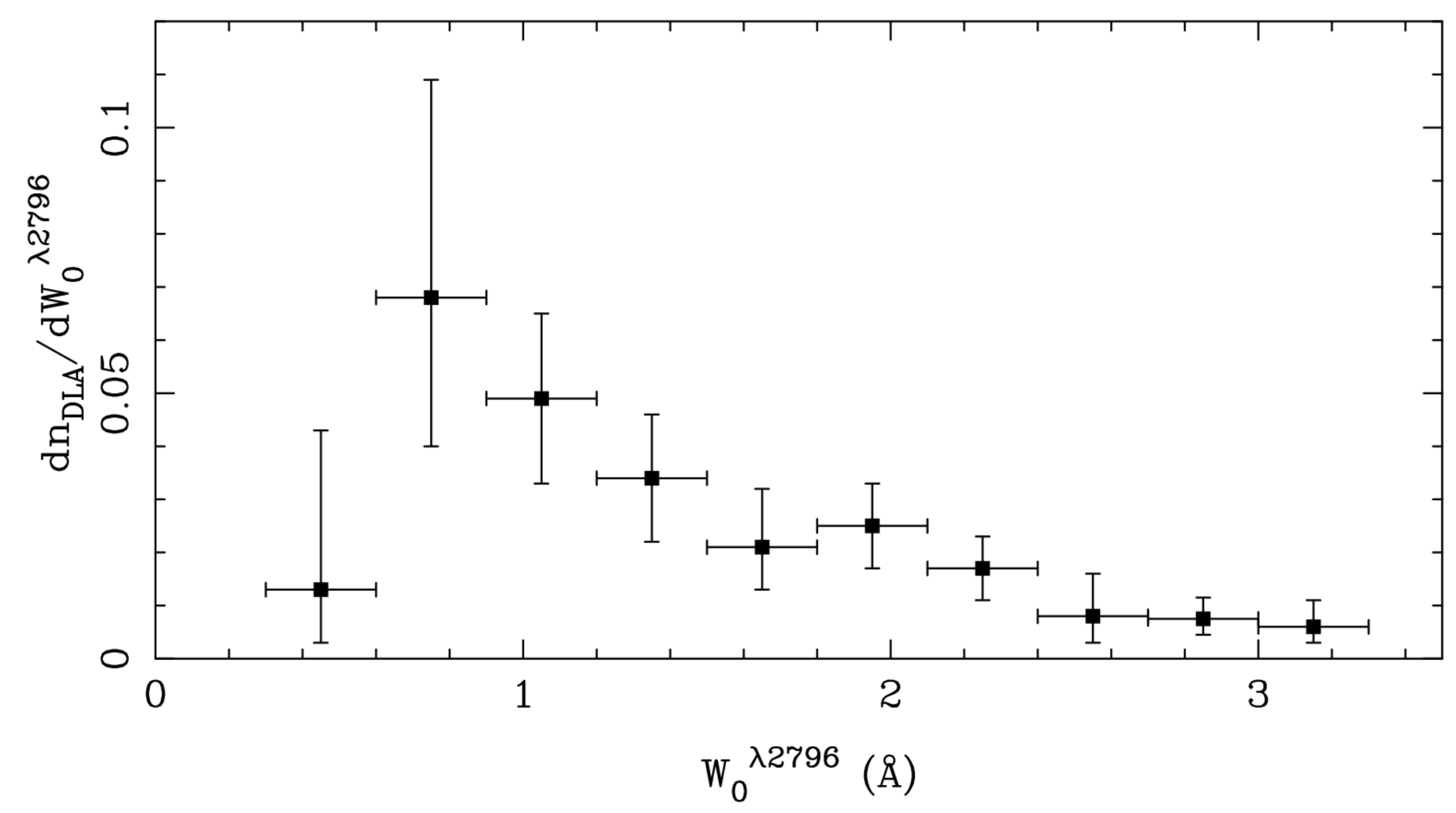}
\caption{The cosmic number density of DLAs as a function of \Wmi, calculated as the product of the two distributions shown in Figure \ref{DLAfracdndW}. \MgII\ systems with 0.6 \AA\ $\le$ \Wmi\ $<$ 1.5 \AA\ contribute 61\% of the total number density of DLAs. }
\label{dnDLAdw}
\end{figure*}

The trend that metallicity increases with \Wmi\ is well known.  Turnshek et al. (2005) used composites of  SDSS spectra binned in \Wmi\ and the mean \HI\ column density for \Wmi$\ge 0.6$ \AA\ from our HST surveys to show the following: 1) that the mean metallicity of \MgII\ absorbers increased with increasing \Wmi, 2) that the \Wmi$=0.6$ \AA\ systems have a mean metallicity of one-hundredth solar ([X/H]$=-2$) and showed little evidence for depletion, and 3) that the amount of depletion increased with increasing \Wmi. P\'eroux et al. (2003), Ledoux et al. (2006), Meiring et al. (2009), M\o ller et al. (2013), Som et al. (2015), Quiret et al. (2016), among  others, have noted a trend between line widths and metallicity, prompting some to infer a mass-metallicity relation for quasar absorbers similar to what is observed for galaxies. With recent results on \MgII\ absorbers suggesting that they arise in bipolar outflows (Bouch\'e et al. 2012; Bordoloi et al. 2014), the mass of the host galaxy may not be the only driver of the velocity width - metallicity relation. In any case, whatever the reason for the correlation, it is recognized that the correlation exists. Does this mean that we are missing the lowest metallicity DLAs since our selection has a cut-off at \Wmi$=0.3$ \AA? 

Since we believe that we have not missed any DLAs through \MgII\ selection (\S4.1), we also have not missed any DLAs with metallicities that are lower than what current samples measure. One way to investigate whether our \MgII-selected DLAs and subDLAs have higher metallicities, is to compare their metallicities to samples of DLAs that were identified serendipitously. Figure \ref{ZvsNHI} shows metallicity measurements of DLAs and subDLAs in \MgII-selected systems and non-\MgII-selected systems at $z<1.65$, i.e., in UV-detected systems. There is now a statistically significant number of non-\MgII-selected DLAs at $z<1.65$ for which metallicity measurements exist in the literature to be able to perform this comparison in a meaningful way. The well-known increase in metallicity with decreasing $N_{\rm HI}$ is observed in this sample as well. However, there is no evidence that the \MgII-selected DLAs have higher metallicities than the non-\MgII-selected ones. A Kolmogorov-Smirnov (KS) test performed on the two samples gives a {\it p}-value of 0.934, indicating that the two samples are drawn from the same parent population at a high level of confidence.

In Figure \ref{XvsW}, we plot the metallicites of DLAs and subDLAs at $z<1.65$ as a function of \Wmi. Both MgII-selected as well as non-MgII-selected systems are included. (There are a few serendipitously discovered DLAs with measurements of \Wmi.)  The trend of increasing metallicity with \Wmi\ is clear, particularly for the DLAs. There are two points to note from this plot. First, since we have shown that there are no DLAs with \Wmi$< 0.3$ \AA, the trend of metallicity with \Wmi\ implies that there must be a floor for DLA metallicities at [X/H] $\sim-2$.  The reason for the metallicity-kinematics relation is far from understood (\S1), but our investigation of neutral gas properties has revealed an apparent floor for both metallicity as well as velocity spread of neutral gas structures at low redshift. It is also interesting to note that studies of the incidence of \Wmi\ show a change in slope for the distribution of $dn/d$\Wmi\ at \Wmi$\sim 0.3$ \AA\ (NTR05; Narayanan et al. 2007), indicating the possibility that this threshold might represent two distinct populations of absorbers.

Second, Figure \ref{XvsW} also shows that measurements of DLA metallicities have apparently sampled \Wmi\ values fairly uniformly across the range of rest equivalent widths 0.3 \AA\ $\le$ \Wmi\ $\le 3.8$ \AA. However, the rest equivalent distribution of \MgII$\lambda2796$ is a strong function of \Wmi, with \Wmi$=4$ \AA\ systems being nearly three orders of magnitude less common than \Wmi$=0.3$ \AA\ systems (NTR05).  Therefore, in the end, observers studying the average cosmic metallicity need to be aware of the fact that by selecting a DLA system with a higher gas velocity spread (i.e., a higher \Wmi\ value) they will be tending to select a DLA with higher metallicity. Thus, an unbiased sample for a DLA metallicity study should be designed to have a relative distribution of \Wmi\ values that matches that of the true DLA population. But, as we have discussed, in order to increase the probability of finding DLAs, our survey methods have tended to select high \Wmi\ systems. We account for this when we calculate the cosmic incidence of DLAs (\S3.2). Similarly, calculations of cosmic metallicity should account for this as well or they might inadvertently be using a non-representative sample.

Figures \ref{DLAfracdndW} and \ref{dnDLAdw} illustrate this further. In Figure \ref{DLAfracdndW}, the left vertical axis and the black solid data points are the same as in Figure \ref{DLAfraction}. On the right vertical axis and as blue data points, we plot the \MgII\ rest equivalent distribution from NTR05, which is plotted for the mean redshift of our sample, $\left<z\right>=0.8$, and for the range of \Wmi\ in our sample, 0.3 \AA\ $\le$ \Wmi\ $< 3.3$ \AA. Figure \ref{dnDLAdw} shows the product of these two quantities, which gives the incidence of DLAs as a function of \Wmi;  \MgII\ systems with 0.6 \AA\ $\le$ \Wmi\ $<$ 1.5 \AA\ are seen to contribute $\sim61$\% of the total number density of DLAs.  Thus, when determining the mean neutral-gas cosmic metallicity of the universe, samples of DLAs should be selected according to the appropriate relative numbers of their \Wmi\ values (or equivalently, velocity spread values). We also point out that the two-orders-of-magnitude spread in metallicity measurements of DLAs at all redshifts is, at least in part, a manifestation of the metallicity-velocity spread relation that exists within absorber samples. That the mean metallicity of DLA absorbers evolves with redshift might then be closely linked to the redshift evolution of the \MgII\ rest equivalent distribution, $dn/d$\Wmi. However, since the \Wmi-metallicity-redshift relation is not the main purpose of this paper, we will not explore this further here.

\section{Conclusions}
We have used the \MgII\ selection method to perform an unbiased survey for DLA absorption line systems at $0.11 < z < 1.65$. The results represent an update to our previous findings reported in RTN06. In particular,  three new subsamples of MgII absorbers were investigated using UV spectroscopy to determine if they had a DLA line indicative of $N_{\rm HI} \ge 2\times10^{20}$ atoms cm$^{-2}$. The previous RTN06 sample had 41 DLAs in 197 \MgII\ systems, and updates to this sample now include 26 DLAs in 96 \MgII\ systems from our HST ACS Prism Survey, three DLAs in 60 \MgII\ systems from our GALEX Archival Survey, and zero DLAs in 16 \MgII\ systems from our MMT-HST COS Survey (although two systems from this survey that did not qualify to be in our statistical sample were DLAs, \S2.3). Each one of these new subsamples had a specific \MgII$\lambda 2796$ rest equivalent width, \Wmi, selection criterion threshold. These thresholds had to be taken into account when deriving the updated DLA statistical results since, as described in \S3.1, the probability of detecting a DLA in a \MgII\ system depends on \Wmi (Figure \ref{DLAfraction}). In total this new, updated sample includes 70 DLAs in 369 \MgII\ systems with \Wmi\ $\ge 0.3$ \AA. Analysis of this new, updated sample indicate the following:

\begin{itemize}

\item[1.] The incidence of DLAs, or product of their gas cross section and their comoving number density, can be described by $n_{\rm DLA}(z) = (0.027 \pm 0.007) (1+z)^{(1.682 \pm 0.200)}$. Only the results of N12 for redshifts $2<z<3.5$ and our current results for $0.11<z<1.65$ were used to derive this relation. It is shown in Figure \ref{dndtfits} as the black curve. This relation appears to hold over the redshift interval $0 < z < 5$ and is in good agreement with the 21 cm emission results reported by B12 at $z=0$. 

\item[2.] The cosmic mass density of neutral gas can be described by $\Omega_{\rm DLA}(z) = (4.77 \pm 1.60)\times10^{-4} (1 + z)^{(0.64\pm 0.27)}$. Once again, only the N12 high redshift and our current low redshift results were used to derive this relation. The curve is shown in Figure \ref{Omtfit}. Given the large uncertainties in our low redshift data points, the fit should not be used to draw any conclusions regarding its extrapolated value to $z=0$, although it is formally more consistent with the Z05 result. 

\item[3.] The results on the HI column density distribution, $f(N)$, are shown in Figure \ref{fNplot}. The best-fitting power law for our data points in the range $20.3 \le \log N_{\rm HI} < 21.8$ has a power law index of $-1.46 \pm 0.20$. At the lowest $N_{\rm HI}$ values our results lie between the $z=0$ 21 cm emission results of Z05 and B12, with the B12 results being several times lower than the Z05 results. In terms of a possible trend, the low $N_{\rm HI}$ tail of the DLA $f(N)$ distribution is seen to decrease as one progresses from high redshift (N12), to low redshift (this study), to $z=0$ (B12), although to establish this we would need to more clearly understand why the results of Z05 and B12 disagree. At the highest $N_{\rm HI}$ values our results are more consistent with the $z=0$ 21 cm emission results of B12, which are also similar to the results of N12. 

\end{itemize}

Finally, in \S4 we addressed the possibilities that the \MgII\ selection method for DLAs leads to either $\left< N_{\rm HI} \right >$ values that are biased high or a metallicity bias. The results shown in Figures \ref{NHIvsWDLAs} and \ref{ZvsNHI}, respectively, show that both of these possibilities can be rejected with certainty in the redshift interval $0.11<z<1.65$, which is the UV DLA redshift regime.  Thus,  at least at $z<1.65$, DLAs found through \MgII\ selection statistically represent the true population of DLAs. We also caution that studies of DLA metallicities should take into the account the relative incidence of DLAs with respect to \Wmi\ (or gas velocity spread) in order to correctly measure the mean cosmic metallicity of the universe.

\section*{Acknowledgments}

We thank Dan Nestor for his contributions to earlier parts of this study. 

The SDSS is managed by the Astrophysical Research Consortium for the Participating Institutions. The Participating Institutions are the American Museum of Natural History, Astrophysical Institute Potsdam, University of Basel, University of Cambridge, Case Western Reserve University, University of Chicago, Drexel University, Fermilab, the Institute for Advanced Study, the Japan Participation Group, Johns Hopkins University, the Joint Institute for Nuclear Astrophysics, the Kavli Institute for Particle Astrophysics and Cosmology, the Korean Scientist Group, the Chinese Academy of Sciences (LAMOST), Los Alamos National Laboratory, the Max-Planck-Institute for Astronomy (MPIA), the Max-Planck-Institute for Astrophysics (MPA), New Mexico State University, Ohio State University, University of Pittsburgh, University of Portsmouth, Princeton University, the United States Naval Observatory, and the University of Washington.

{}

\bsp	
\label{lastpage}
\end{document}